\documentclass[superscriptaddress,nofootinbib,twocolumn,letterpaper,prb]{revtex4-1}

\usepackage{amsmath,amssymb,amsfonts}
\usepackage{mathrsfs}
\usepackage{graphicx}
\usepackage[usenames,dvipsnames]{color}
\usepackage{hyperref}
\usepackage{url}
\usepackage[utf8]{inputenc}
\usepackage{slashed}
\usepackage{subfigure}
\usepackage{slashed,bbm}
\usepackage{graphics,psfrag,epsfig}
\usepackage{dsfont}
\usepackage{setspace}
\newcommand{\beq} {\begin{equation}}
\newcommand{\eeq} {\end{equation}}
\newcommand{\bea} {\begin{eqnarray}}
\newcommand{\eea} {\end{eqnarray}}
\makeatletter
\newcommand*{\toccontents}{\@starttoc{toc}}
\makeatother

\usepackage{hyperref}
\usepackage{xcolor}
\definecolor{dark-red}{rgb}{0.4,0.15,0.15}
\definecolor{dark-blue}{rgb}{0.15,0.15,0.4}
\definecolor{medium-blue}{rgb}{0,0,0.5}

\DeclareMathOperator*{\SumInt}{%
\mathchoice%
  {\ooalign{$\displaystyle\sum$\cr\hidewidth$\displaystyle\int$\hidewidth\cr}}
  {\ooalign{\raisebox{.14\height}{\scalebox{.7}{$\textstyle\sum$}}\cr\hidewidth$\textstyle\int$\hidewidth\cr}}
  {\ooalign{\raisebox{.2\height}{\scalebox{.6}{$\scriptstyle\sum$}}\cr$\scriptstyle\int$\cr}}
  {\ooalign{\raisebox{.2\height}{\scalebox{.6}{$\scriptstyle\sum$}}\cr$\scriptstyle\int$\cr}}
}

\allowdisplaybreaks
\begin{document}

\title{Competing orders at higher-order Van Hove points}

\author{Laura Classen}
\affiliation{School of Physics and Astronomy, University of Minnesota, Minneapolis, MN 55455, USA}
\affiliation{Condensed Matter Physics and Materials Science Department, Brookhaven National Laboratory, Upton, NY 11973, USA}

\author{Andrey V. Chubukov}
\affiliation{School of Physics and Astronomy, University of Minnesota, Minneapolis, MN 55455, USA}

\author{Carsten Honerkamp}
\affiliation{Institute for Theoretical Solid State Physics, RWTH Aachen University, 52074 Aachen, Germany}
\affiliation{JARA-FIT, J\"ulich Aachen Research Alliance - Fundamentals of Future Information Technology, Germany}

\author{Michael M. Scherer}
\affiliation{Institut f\"ur Theoretische Physik, Universit\"at zu K\"oln, 50937 Cologne, Germany}

\date{\today}

\begin{abstract}
Van Hove points are special points in the energy dispersion, where the density of states exhibits analytic singularities.
When a Van Hove point is close to the Fermi level, tendencies towards density wave orders, Pomeranchuk orders, and superconductivity can all be enhanced, often in more than one channel, leading to a competition between different orders and unconventional ground states.
Here we consider the effects from higher-order Van Hove points, around which the dispersion is flatter than near a conventional Van Hove point, and the density of states has a power-law divergence.
We argue that such points are present in intercalated graphene and other materials.
We use an effective low-energy model for electrons near higher-order Van Hove points and analyze the competition between different ordering tendencies using an unbiased renormalization group approach.
For purely repulsive interactions, we find that two key competitors are ferromagnetism and chiral superconductivity.
For a small attractive exchange interaction, we find a new type of spin Pomeranchuk order, in which the spin order parameter winds around the Fermi surface.
The supermetal state, predicted for a single higher-order Van Hove point, is an unstable fixed point in our case.
\end{abstract}
\maketitle

\section{Introduction}

The competition between different types of ordering tendencies plays a key role in many quantum materials.
For example, unconventional superconductivity often develops near a charge or spin order and is viewed as mediated by soft charge or spin fluctuations.
Within an itinerant electron scenario, the formation of an ordered phase can be understood as an instability of the parent electron liquid, driven by excitations around the Fermi energy.
Therefore, the density of states (DOS) near the Fermi level and the geometry of the Fermi surface strongly affect the low-energy phase formation.
In a 2D crystal, both quantities can change significantly when the fermionic dispersion possesses a saddle point, which is one of the prominent examples of Van Hove points~\cite{PhysRev.89.1189}.
The DOS near such a point diverges logarithmically and the Fermi surface transforms between a hole- and electron-like form.
If the Fermi level lies in the vicinity of a Van Hove point, the singular DOS determines the physical behavior due to the large number of available low-energy states.
In particular, interaction effects get amplified not only in the particle-particle, but also in the particle-hole channels, leading to the notion of competing orders.
A prototypical example is the interplay of spin-density-wave order and $d$-wave superconductivity near Van Hove filling in the Hubbard model on the square lattice~\cite{HUR20091452}.

For electrons on the honeycomb lattice, e.g., in single-layer graphene, the competition is again between $d$-wave superconductivity and spin-density-wave order, but the ordered states are more non-trivial:
$d$-wave superconductivity is  chiral~\cite{nandkishore2012chiral,PhysRevB.86.020507,PhysRevB.85.035414,nandkishore2014superconductivity,black2014chiral}, and spin-density-wave order is a half-metal~\cite{PhysRevLett.108.227204}, which additionally breaks lattice translational symmetry~\cite{PhysRevB.86.115443}.
On the other hand, while for square-lattice systems the Van Hove points are located reasonably close to the Fermi level already at charge neutrality, they are at higher energies for electrons on the honeycomb lattice, and it requires a substantial amount of doping to reach them.
Recently, such doping levels have been made accessible by Gadolinium intercalation of graphene~\cite{PhysRevB.100.121407}.
The intercalation leads to a renormalization of the band structure, which reduces the bandwidth and, hence, the value of the chemical potential required to bring the Van Hove points to the Fermi level.
However, the intercalation also brings another effect: it flattens the band dispersion around the Van Hove points.
This flattening gives rise to a stronger power-law singularity of the DOS, which can qualitatively affect the balance between different ordering tendencies.
In particular, it suppresses finite wave-vector density-wave fluctuations and enhances fluctuations with zero momentum transfer, e.g., Stoner-type instabilities.
Consequently, a new type of competition occurs between the pairing and zero-momentum instabilities in the particle-hole channel.

A Van Hove point with a power-law divergence of the DOS has been termed higher-order Van Hove (HOVH) point, as opposed to a conventional Van Hove (CVH) point.
HOVH points were proposed to exist in moir\'e superlattices, e.g., twisted bilayer graphene and trilayer graphene, in which the twist angle, pressure, or an electric field can be used to tune the band structure~\cite{yuan2019magic}.
Germanene on MoS$_2$ shows similar effects as intercalated graphene: a reduced Van Hove energy and the band flattening around the Van Hove points \cite{PhysRevB.99.201106}.
Other examples
for systems with HOVH points
 include biased bilayer graphene at charge neutrality \cite{PhysRevB.95.035137} and magnetic-field-tuned Sr$_3$Ru$_2$O$_7$\cite{PhysRevLett.123.207202}  and $\beta$-YbAlB$_4$\cite{PhysRevLett.109.176404}.
The case of a single HOVH point in the Brillouin zone has recently been studied in Ref.~\onlinecite{PhysRevResearch.1.033206}.
It was shown that fluctuations around this single HOVH point drive the system towards a critical non-Fermi-liquid ground state, dubbed a supermetal.

In this work, we analyze the competition of ordering tendencies arising from the presence of {\it  multiple} HOVH points near the Fermi level.
In this more general case, additional types of couplings occur, and we show that the supermetal state becomes an unstable fixed point.
Instead, the system develops an instability towards either superconductivity, or
Pomeranchuk order.

We set up a renormalization group (RG) framework within an effective low-energy model for electrons near the HOVH points, with parameters appropriate for Gadolinium-doped graphene.
This allows us to account for the interplay between different ordering tendencies and identify the leading instability.
We show that chiral superconductivity can still develop, as for the case of a CVH point, but the pair-hopping term, which drives it, needs to be sufficiently strong compared to other couplings.
For other ranges of interactions, we find a ferromagnetic instability and a special $d$-wave spin Pomeranchuk order, in which the spin order parameter winds around the Fermi surface.

\section{Higher-order Van Hove singularity in graphene}

It was shown in Ref.~\onlinecite{PhysRevB.100.121407} that the doping levels needed to reach the Van Hove energy in graphene can be made accessible by intercalation, with large-scale homogeneity and very good crystallinity.
In the process, the electronic spectrum undergoes strong renormalizations, which not only bring down the Van Hove energy, but also flatten the energy dispersion around the M points, i.e., transform CVH points into HOVH points.
While electronic correlations may be responsible for the band renormalization~\cite{PhysRevB.100.121407}, we can model their effect by introducing an effective single-particle Hamiltonian for electrons on the honeycomb lattice with hopping up to the third neighbor.
This allows us to qualitatively reproduce the observed band flattening along the K-M direction and the measured Fermi surface geometry.
However, we emphasize that our analysis of the competing orders below does not depend on the precise band structure  or the mechanisms causing it.

\subsection{Effective hopping Hamiltonian}

\begin{figure}[t!]
\begin{center}
\includegraphics[width=\columnwidth]{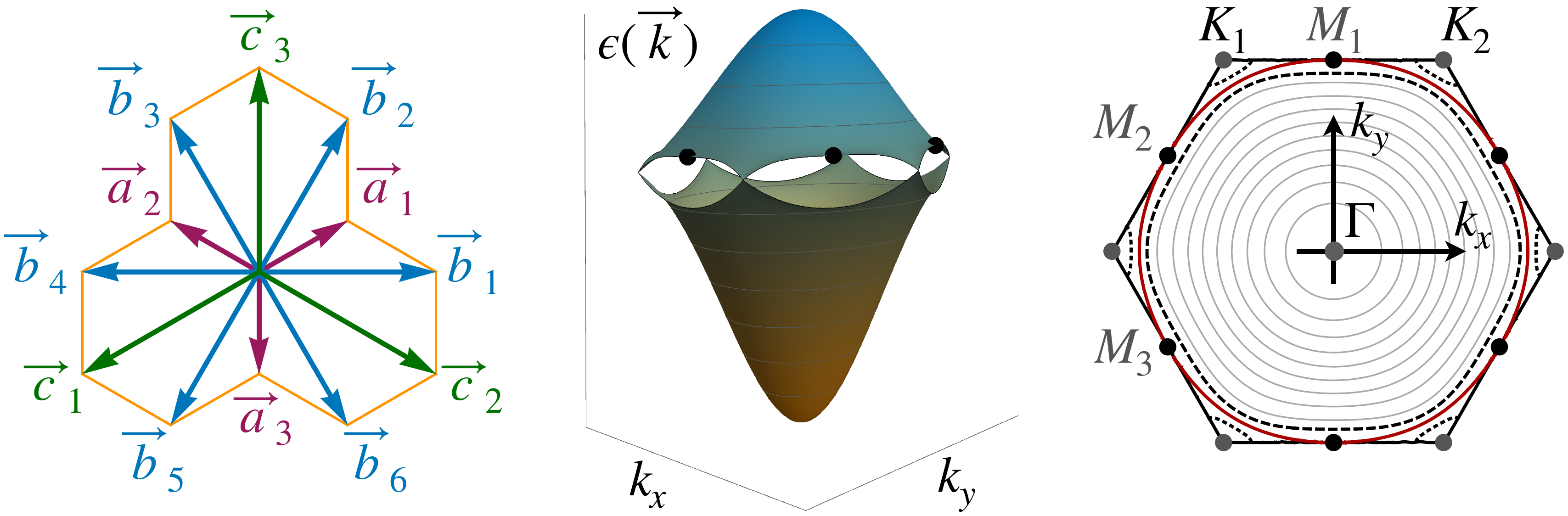}
\end{center}
\caption{\textbf{Lattice and tight-binding model.}
\textit{Left panel:} Lattice in real space with neighboring vectors $\vec a_n, \vec b_n, \vec c_n$. The nearest-neighbor vectors on the honeycomb lattice are $\vec{a}_1=(\sqrt{3},1)/2$, $\vec{a}_2=(-\sqrt{3},1)/2$, $\vec{a}_3=(0,-1)$, the second-nearest-neighbor vectors are $\vec{b}_1=(\sqrt{3},0), \vec{b}_2=(\sqrt{3},3)/2, \vec{b}_3=(-\sqrt{3},3)/2$, $\vec{b}_4=-\vec{b}_1, \vec{b}_5=-\vec{b}_2, \vec{b}_6=-\vec{b}_3$,  and the third-nearest-neighbor vectors are $\vec{c}_1=-2\vec{a}_1, \vec{c}_2=-2\vec{a}_2, \vec{c}_3=-2\vec{a}_3$.
\textit{Middle panel:} Energy dispersion for $t_1=1, t_2=0.1, t_3=0.2$. The valence and conduction band touch at the Dirac points. The Van-Hove points appear at the $M$ points, which are marked by black dots for the conduction band.
\textit{Right panel:} Energy contours for the same hopping amplitudes and the high-symmetry points $K_1=2\pi/3(1/\sqrt{3},1), K_2=2\pi/3(-1/\sqrt{3},1)$ and $M_1=\pi(0,2/3)$, $M_2=\pi(-1/\sqrt{3},1/3)$, $M_3=-\pi(1/\sqrt{3},1/3)$. The Fermi level at Van-Hove filling is given by the red line. At this filling, the system undergoes a Lifshitz transition from a closed to an open Fermi surface. For a small variation of the filling, the Fermi surface is either closed as demonstrated by the nearby dashed line, or changes to open Fermi-surface pockets given by the dotted lines around the $K$ points.}
\label{fig:reallattice}
\end{figure}

The effective Hamiltonian including up to third-neighbor hopping reads
\begin{align}
	H_0=&\Big[t_1\sum_{\langle i,j\rangle, \sigma}c_{i\sigma}^\dagger c_{j\sigma}+t_2\sum_{\langle\!\langle i,j\rangle\!\rangle, \sigma}c_{i\sigma}^\dagger c_{j\sigma}\nonumber\\
	&+t_3\sum_{\langle\!\langle\!\langle i,j\rangle\!\rangle\!\rangle, \sigma}c_{i\sigma}^\dagger c_{j\sigma}+\mathrm{H.c.}\Big]-\mu \sum_{i\sigma} n_{i\sigma}\,.
	\label{eq:model}
\end{align}
We have introduced $c^{(\dagger)}_{i,\sigma}$ as the fermion annihilation (creation) operators at site $i$ and spin projection $\sigma \in \{\uparrow, \downarrow\}$. The nearest-, second-nearest and third-nearest neighbor hopping amplitudes are $t_1, t_2$ and $t_3$, and the Fermi level can be adjusted with the chemical potential $\mu$. We have defined $n_{i,\sigma}=c_{i\sigma}^\dagger c_{i\sigma}$ as the particle number operator. The honeycomb lattice and the locations of first, second, and third neighbors are sketched in Fig.~\ref{fig:reallattice}.

From the model Eq.~\eqref{eq:model} we obtain the energy bands
\begin{align}
	\epsilon_\pm(\vec{k})=\pm \left| t_1\alpha(\vec{k})+t_3\gamma(\vec{k})\right| -t_2\beta(\vec{k}) -\mu \,.
	\label{eq:dispersion0}
\end{align}
with
$\alpha(\vec{k})=\sum_{n=1}^3 e^{-i\vec{k}\cdot\vec{a}_n}$,
$\beta(\vec{k})=\sum_{n=1}^6 e^{-i\vec{k}\cdot\vec{b}_n}$,
and
$\gamma(\vec{k})=\sum_{n=1}^3 e^{-i\vec{k}\cdot\vec{c}_n}$, where $\vec{a}_n,\vec{b}_n,\vec{c}_n$ denote nearest-, second- and third-neighbor vectors, see Fig.~\ref{fig:reallattice}.

The energy dispersion and the corresponding Fermi surface depend on the choice of the hopping amplitudes $t_1,t_2,t_3$ and the chemical potential $\mu$.
For definiteness, we consider the branch $\epsilon_+ ({\vec k})$ and discuss how it changes with the chemical potential.
At small $\mu$,  the Fermi surface consists of six pockets around the  Dirac points, see Fig.~\ref{fig:reallattice}.
As $\mu$ increases, the edges of the Fermi pockets come closer to each other, and at
\begin{align}
\mu=t_1+2t_2-3t_3\,,
\end{align}
they merge at the three special high-symmetry points on the edges of the first Brillouin zone, i.e. $M_1,M_2,M_3$, see Fig.~\ref{fig:reallattice}.
At larger $\mu$, the Fermi surface is a closed loop, centered at the $\Gamma$ point.

\begin{figure}[t!]
\begin{center}
\includegraphics[width=0.8\columnwidth]{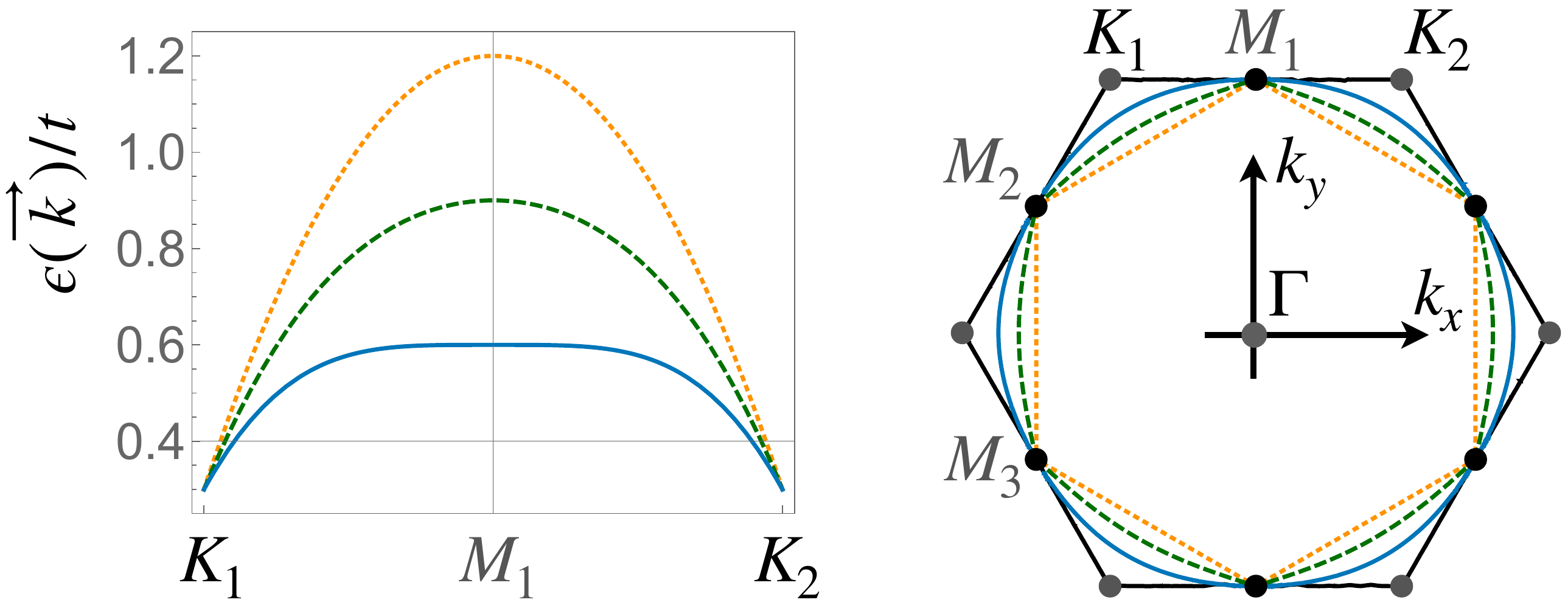}
\end{center}
\caption{\textbf{Band flattening at Van Hove point.}
\textit{Left panel:} Dispersion along $K_1-M_1-K_2$ for fixed $t_2=0.1t_1$ and varying $t_3=0$ (orange, dotted),  $t_3=0.1t_1$ (green, dashed), and $t_3=0.2t_2$ (blue, solid). The last value of $t_3$ leads to a HOVH singularity.
\textit{Right panel:} Corresponding Fermi surface.}
\label{fig:t3effect}
\end{figure}

The points $M_p$, $p\in \{1,2,3\}$, are Van Hove points.
We can verify this by expanding the dispersion around them
\begin{align}\label{eq:m1approx}
	\epsilon_{M_1}(\vec{x})  &= b_1y^2 -a_1x^2 + ...\,,\\
	\epsilon_{M_2}(\vec{x})  &=  a_2 x^2-c_2 x y+ b_2 y^2+...\\
	&=b_2 \left(y-c_2 x/(2b_2)\right)^2 - \frac{1}{b_2}(c^2_2/4 - a_2 b_2) x^2+...\label{eq:m2approx}\,,\nonumber\\
	\epsilon_{M_3}(\vec{x}) &=a_2 x^2-c_2 x y+ b_2 y^2+... \\
	&=b_2 \left(y +c_2 x/(2b_2)\right)^2 - \frac{1}{b_2}(c^2_2/4 - a_2 b_2) x^2+...\label{eq:m3approx}\,,\nonumber
\end{align}
where $\epsilon_{M_p}(x,y)=\epsilon_\pm(M_{p,x}+x,M_{p,y}+y) + \mu$, and the dots denote higher order terms in $x,y$.
The coefficients are given by the hoppings $t_1,t_2,t_3$, see App.~\ref{app:Mexp}.
All $a_p, b_p, c_p \geq 0$ and $c_p \geq (2 a_2 b_2)^{1/2}$.
Since the dispersion is quadratic, with opposite signs along the two directions, the DOS is logarithmically singular.
This holds as long as the prefactors are non-zero, i.e. $a_1, b_1 >0$ for $\epsilon_{M_1}(\vec{x})$ and $b_2,(c_2^2/4-a_2b_2)>0$ for $\epsilon_{M_{2/3}}(\vec{x})$.

\subsection{Higher-order Van Hove points}

The CVH points become HOVH points when one of the prefactors in Eq.~(\ref{eq:m1approx}) vanishes, and one has to expand further to get the dispersion in the corresponding direction.
In our model this happens for
\begin{align}
	t_3 \to t_{3,c}=\left(t_1-2t_2\right)/4\,.
\end{align}
For this special case, $a_1$ and $c^2_2/4 - a_2b_2$ in Eq.~\eqref{eq:m1approx} vanish.
We show the flattening of the dispersion for increasing $t_3$ in Fig.~\ref{fig:t3effect}, together with the change of the Fermi surface, which becomes rounder. This qualitatively mimics the effect observed for gadolinium intercalation in graphene\cite{PhysRevB.100.121407}.

\begin{figure}[t]
\begin{center}
\includegraphics[width=0.95\columnwidth]{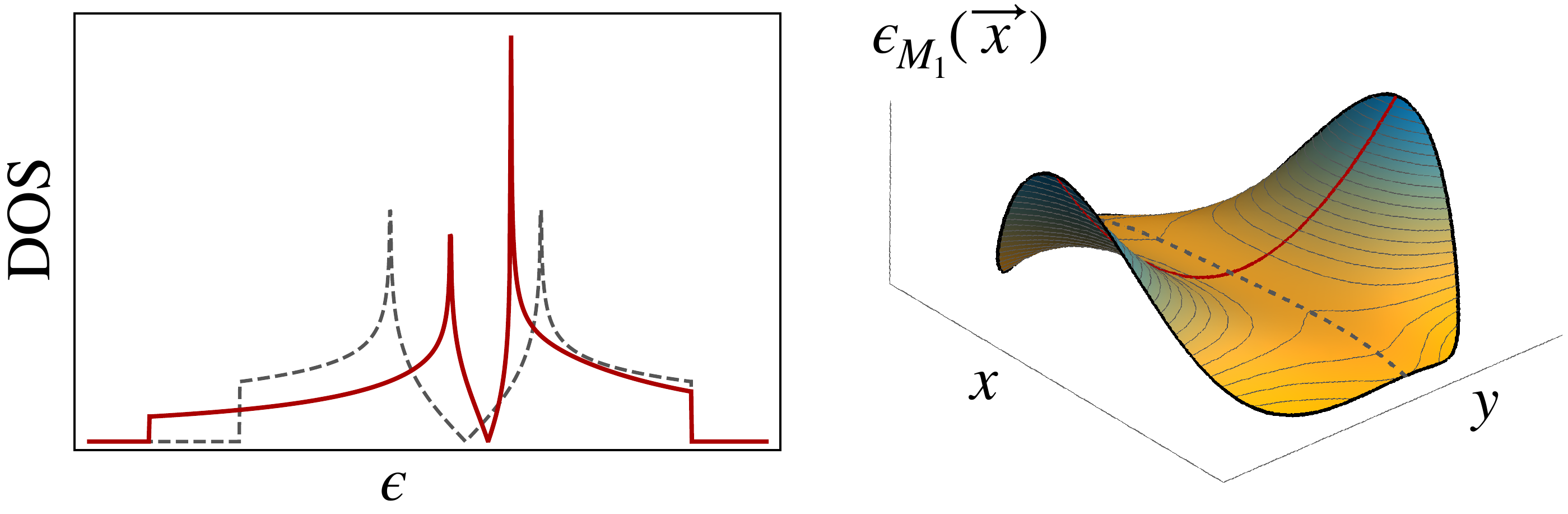}
\end{center}
\caption{\textbf{DOS and saddle point.}
\textit{Left panel:} DOS of the band dispersion for hopping parameters $t_1\!=\!1,t_2\!=\!1/10, t_3\!=\!\frac{1}{4}(1-2/10)$ (solid line) and DOS for hopping parameters $t_1\!=\!1, t_2\!=\!t_3\!=\!0$ for comparison (dashed line).
\textit{Right panel:} Corresponding higher-order saddle point at $M_1$.}
\label{fig:hosp}
\end{figure}

For $t_3=t_{3,c}$, we have to expand to higher order, i.e.
\begin{align}\label{eq:m1approx_1}
	\epsilon_{M_1}(\vec{x})  &= b_1y^2 -d_1x^4 + ...\,,\\
	\epsilon_{M_2}(\vec{x}) &= b_2 (y-c_2 x/(2b_2))^2 - d_2 x^4+...\,,\\
	\epsilon_{M_3}(\vec{x})  &= b_2 (y +c_2x/(2b_2))^2 - d_2 x^4+...\,,
\end{align}
with $d_{1,2}>0$. The saddle-type dispersion near this HOVH point is shown in Fig.~\ref{fig:hosp}. For such a dispersion, the DOS shows a power-law divergence
\bea
	\rho (\epsilon) =
	\begin{cases}
	\rho_+ \epsilon^{-1/4} & {\text{for }} \epsilon > 0, \\
	\rho_- |\epsilon|^{-1/4} & {\text{for }} \epsilon < 0,
	\end{cases}
\label{ch_1}
\eea
where $\rho_+=\Gamma[1/4]/\left[8\pi^{5/2}( b_1^2  d_1)^{1/4}\right]$ and $\rho_-=\rho_+/\sqrt{2}$, cf. Ref.~\onlinecite{PhysRevResearch.1.033206}.
This divergence is stronger than the logarithmic one at a CVH point.
The singular behavior of the DOS near the HOVH point can be determined from a scaling argument~\cite{yuan2019classification}, see Appendix ~\ref{app:Mexp}.

We also consider the generalized case with
\begin{align}\label{eq:m1approx_2}
	\epsilon_{M_1}(\vec{x}) &= b_1y^2 -d_1x^{2\alpha} + ...\,,\\
	\epsilon_{M_2}(\vec{x})  &= b_2 (y-c_2 x/(2b_2))^2 - d_2 x^{2\alpha}+...\,,\\
	\epsilon_{M_3}(\vec{x})  &= b_2 (y +c_2x/(2b_2))^2 - d_2 x^{2\alpha}+...\,,
\end{align}
where $\alpha >1$.
The case $\alpha =1$ corresponds to a CVH point, the case $\alpha =2$ to the HOVH point in our model of intercalated graphene.
For $\alpha<2$, this generalized saddle-point dispersion can also be interpreted to effectively model the case where the system is slightly doped away from a HOVH point.

The DOS for the generalized dispersion in Eq.~\eqref{eq:m1approx_2} is
\bea
	\rho (\epsilon) =
	\begin{cases}
	\rho_+ \epsilon^{-\kappa} & {\text{for }} \epsilon > 0, \\
	\rho_- |\epsilon|^{-\kappa} & {\text{for }} \epsilon < 0,
	\end{cases}
\label{ch_2}
\eea
where $\kappa = 1/2 - 1/(2\alpha)$, $\rho_+=\Gamma[1/(2\alpha)] \Gamma[1/2 -1/(2\alpha)]/\left(4 \alpha \pi^{5/2} b_1^{1/2}  d_1^{1/(2\alpha)}\right)$ and $\rho_-=\rho_+ \sin{(\pi/(2\alpha))}$, cf. Ref.~\onlinecite{PhysRevResearch.1.033206}.
For $\alpha =2$, we recover $\kappa =1/4$.  When $\alpha \to 1$, $\kappa \to 0$, and $\rho_{\pm}$  formally diverges  as $1/\kappa$.
The divergence becomes $(1 - (\epsilon/\Lambda)^\kappa)/\kappa = \ln{\Lambda/\epsilon}$, once we keep a UV cutoff $\Lambda$. The logarithmic divergence is the expected result for a CVH point.
There are other examples of systems with a HOVH singularity with various exponents.
The HOVH singularity in twisted bilayer graphene is also described by $\kappa=1/4$, see Ref.~\onlinecite{yuan2019magic}.
In bilayer graphene, one can tune the dispersion with an interlayer voltage bias to a power-law singularity with $\kappa=1/3$ at charge neutrality\cite{PhysRevB.95.035137}.
Sr$_3$Ru$_2$O$_7$\cite{PhysRevLett.123.207202} and $\beta$-YbAlB$_4$\cite{PhysRevLett.109.176404} are expected to have a HOVH singularity with $\kappa=1/2$.

\section{Patch model}
\label{sec:patch}

Because the DOS has a power-law singularity near the HOVH points, the low-energy physics is determined by fermions with momenta near these points.  Accordingly, we restrict our consideration to momentum states in patches around the HOVH points.
The patch size is related to the UV energy cutoff $\Lambda$.
We assume that degrees of freedom with energies larger than $\Lambda$ are integrated out, and microscopic information is incorporated into the bare parameters of the effective patch model.

\begin{figure}[t!]
\begin{center}
\includegraphics[width=0.9\columnwidth]{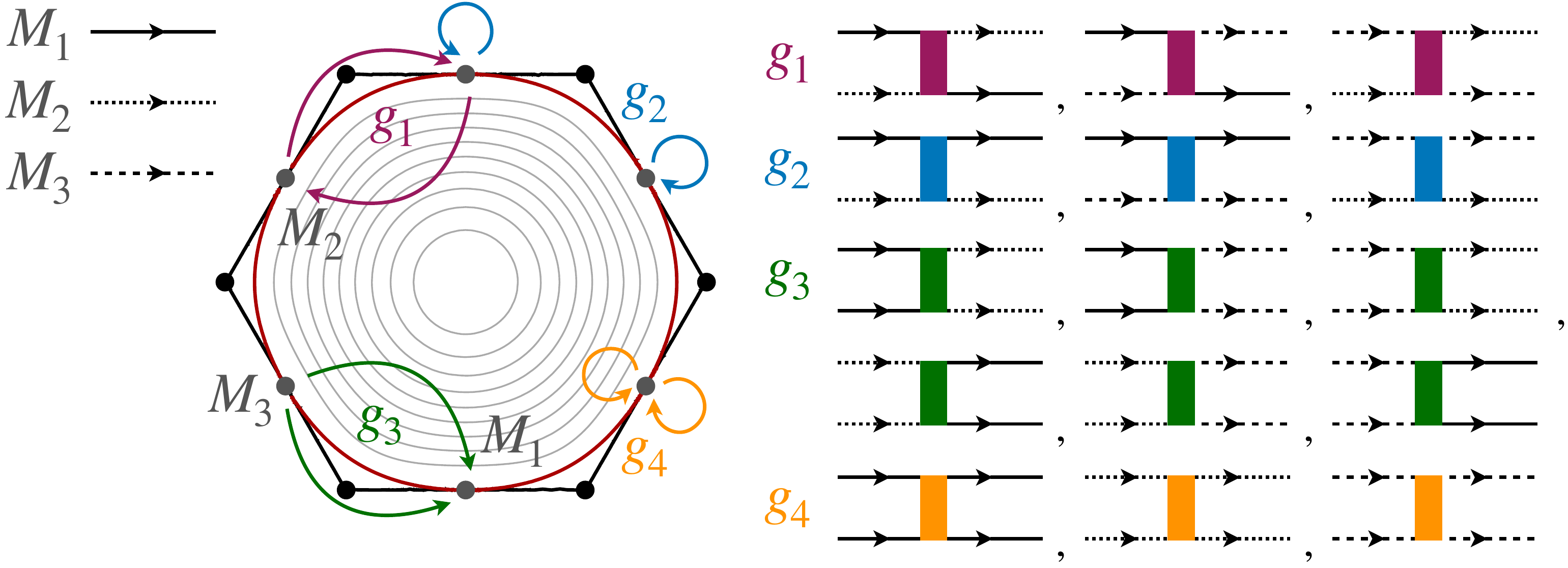}
\end{center}
\caption{\textbf{Three-patch model and interaction couplings.}
Graphic representation of the four interaction couplings $g_i$, $i\in \{1,...,4\}$ representing the scattering processes between the three $M_p$ points for the Van Hove doped dispersion. Solid, dashed and dotted lines represent electrons near the three $M_p$ points.}
\label{fig:gology}
\end{figure}

Within the effective model, we  include all scattering processes between fermions near the HOVH points, allowed by symmetry and momentum conservation.
This gives four different couplings $g_i$, $i\in \{1,...,4\}$, like for the case of CVH points at $M_p$, cf. Ref.~\onlinecite{nandkishore2012chiral}.
The interaction part of the effective Hamiltonian then reads
\begin{align}
H_{g}=\sum_{\substack{k_1\ldots k_3\\\sigma \sigma'}}\sum_{\substack{p,p'=1\\p\neq p'}}^3 \Big[&\ g_1 c_{p'\sigma k_3}^\dagger c_{p \sigma k_4}^\dagger c_{p' \sigma' k_2} c_{p \sigma k_1} \nonumber \\
+ &\ g_2 c_{p \sigma k_3}^\dagger c_{p' \sigma k_4}^\dagger c_{p' \sigma' k_2} c_{p \sigma k_1}\nonumber \\
+ &\ g_3 c_{p' \sigma k_3}^\dagger c_{p' \sigma k_4}^\dagger c_{p \sigma' k_2} c_{p \sigma k_1}\Big] \nonumber \\
+\sum_{\substack{k_1\ldots k_3\\\sigma \sigma'}}\sum_{\substack{p=1}}^3
&\ g_4  c_{p \sigma k_3}^\dagger c_{p \sigma k_4}^\dagger c_{p \sigma' k_2} c_{p \sigma k_1}.
\end{align}
Here, $c_{p\sigma k}$ is the annihilation operator for an electron in the vicinity of the point $M_p$, $p\in \{1,2,3\}$, with momentum $M_p+k$ and spin $\sigma$. The momentum $k$ is restricted to the patch around $M_p$.
The couplings are independent on the flavor index $p$ due to sixfold rotational symmetry.
The scattering processes are sketched in Fig.~\ref{fig:gology}.
We note in passing that an analogous description can be derived for the square lattice with the only difference that there are only two patches $p=1,2$, cf. Ref.~\onlinecite{PhysRevLett.81.3195}.

\section{Susceptibilities}
\label{sec:bubbles}

The interactions receive corrections through different scattering channels.
These corrections grow with decreasing $T$ and, if the dressed interaction diverges at a finite $T$ in at least one channel, the Fermi liquid is not the stable ground state.
In the patch model, potential divergences can occur in the particle-particle or particle-hole channel, due to processes with momentum transfer near zero or near $M_p$.
To understand the relative strength of various corrections, we first compute the corresponding particle-particle and particle-hole susceptibilities for free fermions, i.e.
\begin{align}
	\label{eq:chipp}
	\chi_{\mathrm{pp}}^X&:= T\sum_\omega \int \frac{d^2k}{(2\pi)^2}\, G_0(\omega,k)G_0(-\omega,X-k)\,,\\
	\chi_{\mathrm{ph}}^X&:= -T\sum_\omega \int \frac{d^2k}{(2\pi)^2}\, G_0(\omega,k)G_0(\omega,X+k)\,,
	\label{eq:chiph}
\end{align}
where  $G_0(\omega, q)=1/[i\omega-\epsilon(q)]$ and the wave-vector $X$ is either $0$ or $M_p$. We have set the frequency transfer to zero because there the corrections are the largest.

For $\kappa = 0$, i.e. the case of a CVH singularity, $\chi_{\mathrm{pp}}^0$  diverges as  $\ln^2\Lambda/T$, and $\chi_{\mathrm{ph}}^{M_{p}}$ either diverges as $\ln^2\Lambda/T$  for a nested Fermi surface with nesting vector $M_p$, or as $\ln\Lambda/T$ for non-perfect nesting\cite{nandkishore2012chiral,PhysRevLett.81.3195}.
The susceptibilities $\chi_{\mathrm{pp}}^{M_p}$ and $\chi_{\mathrm{ph}}^0$ diverge less strongly, as $\ln\Lambda/T$ even for perfect nesting.
Then the thermal evolution of the couplings comes primarily from renormalizations in the particle-particle channel at zero momentum transfer and in the particle-hole channel at momentum transfer $M_p$, leading to a competition between tendencies towards a spin-density wave order and superconductivity.
The situation changes  qualitatively at a HOVH point, where the DOS diverges with a power law.
We will show below that in this case  $\chi_{\mathrm{pp}}^0$ and $\chi_{\mathrm{ph}}^0$ diverge as $1/T^\kappa$, while $\chi_{\mathrm{pp}}^{M_p}$ remains logarithmically singular, and $\chi_{\mathrm{ph}}^{M_p}$  becomes constant.
In this case, the key ordering tendencies  are superconductivity  and $q=0$ spin and charge orders.

\subsection{Zero momentum transfer}

The particle-hole susceptibility with zero momentum transfer is
\begin{align}
\chi_{\mathrm{ph}}^0&=-T\sum_\omega\int \frac{d^2k}{(2\pi)^2} \frac{1}{[i\omega-\epsilon_M(k)]^2} \nonumber\\
&= -T\sum_\omega \int d\epsilon \rho(\epsilon) \frac{\partial}{\partial \epsilon}\frac{1}{i\omega-\epsilon} = -\int d\epsilon \rho(\epsilon) \frac{\partial}{\partial \epsilon} n_F(\epsilon)\nonumber\\
&= \frac{1}{4T}\int d\epsilon \frac{\rho_0}{|\epsilon|^\kappa}\frac{1}{\cosh^2(\epsilon/2T)}\nonumber \\
&=\frac{\rho_0}{T^{\kappa}} f(\kappa)\,,\label{eq:chiph0}
\end{align}
where  $n_F(\epsilon)$ is the Fermi function, $\rho_0=(\rho_++\rho_-)/2$, and we defined $f(\kappa)= \frac{1}{4}\int d\epsilon\, |\epsilon|^{-\kappa}\cosh^{-2}(\epsilon/2)$.
In the limit $\kappa\rightarrow 1/4$, we obtain $f (\kappa =1/4) \approx 1.08$.
For $\kappa\rightarrow 0$, the DOS becomes a logarithmic function and we recover the logarithmic temperature dependence in $\chi_{\mathrm{ph}}^0$.
We see that for $\kappa >0$, $\chi_{\mathrm{ph}}^0$ increases by a power-law as $T$ decreases.

For the particle-particle susceptibility $\chi_{\mathrm{pp}}^0$, using inversion symmetry $\epsilon(k)=\epsilon(-k)$, we obtain
\begin{align}
	\chi_{\mathrm{pp}}^0&=T\sum_\omega\int \frac{d^2k}{(2\pi)^2} \frac{1}{[i\omega-\epsilon_M(k)][-i\omega-\epsilon_M(-k)]} \nonumber\\
&=- \int d\epsilon \rho(\epsilon) \frac{n_F(\epsilon)-n_F(-\epsilon)}{2\epsilon}
  \nonumber\\
&=\frac{\rho_0}{T^{\kappa}} g(\kappa)\,,
\label{eq:chipp0}
\end{align}
where $g(\kappa)=\frac{1}{2}\int d\epsilon\, |\epsilon|^{-(1+\kappa)}|\tanh(\epsilon/2)|$.
For $\kappa =1/4$, $g(1/4)\approx 4.33$, for $\kappa \to 0$, $g (\kappa \to 0) \propto 1/\kappa$.
Combining the last behavior with the logarithmic divergence of the DOS in this limit, we find $\chi_{\mathrm{pp}}^0 \propto (\ln{\Lambda/T})^2$, as expected for a CVH point.

\subsection{Finite momentum transfer}

In contrast to $\chi_{\mathrm{pp/ph}}^0$, the susceptibilities at the momentum transfer $M_p$ do not exhibit a power-law divergence.
For definiteness, we consider $\chi_{\mathrm{pp/ph}}^{M_1}$.  For the particle-hole susceptibility we obtain
\begin{align}
\hspace{-0.1cm}
\chi_{\mathrm{ph}}^{M_1}= &-T\sum_\omega\int\!\! \frac{d^2k}{(2\pi)^2} \frac{1}{[i\omega-\epsilon_{M_3}(k)][(i\omega-\epsilon_{M_2}(k)]}\nonumber\\
=&-\int\!\!\frac{d^2k}{(2\pi)^2} \frac{n_F(\epsilon_{M_3}(k))-n_F(\epsilon_{M_2}(k))}{\epsilon_{M_3}(k) - \epsilon_{M_2}(k)}\nonumber\\
\approx &\frac{1}{2\tilde c_2}\!\int^{\frac{\Lambda}{T}}\!\!\frac{d^2\tilde k}{(2\pi)^2}\frac{\frac{1}{\tilde k_x \tilde k_y}\sinh(\tilde c_2\tilde k_x \tilde k_y)}{\cosh(\tilde k_x^2 + \tilde k_y^2)+\cosh(\tilde c_2\tilde k_x \tilde k_y)}.
\label{ch_3}
\end{align}
where we have rescaled $k_x=\sqrt{T/a_2}\tilde k_i,k_y=\sqrt{T/b_2}\tilde k_y$ and introduced $\tilde c_2={c_2}/{\sqrt{a_2 b_2}}$.
For the case of a pure HOVH point, $\tilde c_2 =2$.
In this case, $\chi_{\mathrm{ph}}^{M_1}$ remains finite.
Indeed, a potential singular temperature dependence in Eq.~\eqref{ch_3} can come from the singularity at the upper limit of the integration over $d^2 {\tilde k}$  for $\Lambda/T\to\infty$.
Using polar coordinates, we can re-express the potential singularity in Eq.~\eqref{ch_3} as
\beq
\int^{\frac{\Lambda}{T}} \frac{dr}{r} \int_{\sin{(2\phi)} > 2/{\tilde c}} \frac{d \phi}{\sin{(2\phi)}} \frac{e^{r^2 (\frac{\tilde c}{2} \sin{(2\phi)}-1)}}{1 + e^{r^2 (\frac{\tilde c}{2}\sin{(2\phi)}-1)}}\,.
\label{ch_5}
\eeq
In case $\tilde c_2 =2$, the integration over $\phi$ gives $1/r^2$, and the integral over $r$ converges, i.e.
\beq
\chi^{M_1}_{\mathrm{ph}}\rightarrow \text{const.}
\eeq
For a quadratic dispersion along the $x$~direction ${\tilde c}_2 > 2$.
In this case, there is a finite range of angles~$\phi$, for which $({\tilde c}/2) \sin{\phi} >1$.
In this range, the integration over $\phi$ now yields a finite number, and the integral over $r$ gives $\ln{\Lambda/T}$.  This is the expected behavior for a CVH point.

For  $\chi_{\mathrm{pp}}^M$, we obtain
\begin{align}
\hspace{-0.1cm}
\chi_{\mathrm{pp}}^{M_1}=&-T\sum_\omega\int\!\! \frac{d^2k}{(2\pi)^2} \frac{1}{[i\omega-\epsilon_{M_3}(k)][(i\omega+\epsilon_{M_2}(k)]}\nonumber\\
=&-\int\!\!\frac{d^2k}{(2\pi)^2}  \frac{n_F(\epsilon_{M_3}(k))-n_F(-\epsilon_{M_2}(k))}{\epsilon_{M_3}(k) + \epsilon_{M_2}(k)}\nonumber\\
\approx &\frac{1}{2}\int^{\frac{\Lambda}{T}}\!\!\frac{d^2\tilde k}{(2\pi)^2}\frac{\frac{\sqrt{a_2b_2}}{\tilde k_x^2 +\tilde k_y^2}\sinh(\tilde k_x^2 +\tilde k_y^2)}{\cosh(\tilde k_x^2 + \tilde k_y^2)+\cosh(\tilde c_2\tilde k_x \tilde k_y)}\,.
\end{align}
Using polar coordinates, we find that
\beq
\chi_{\mathrm{pp}}^{M_1} \propto \ln \frac{\Lambda}{T}\,.
\eeq
This result holds for all $\kappa$.
To verify the expressions for the susceptibilities, we computed $\chi_{\mathrm{ph/pp}}^0$ and $\chi_{\mathrm{ph/pp}}^M$ numerically, by integrating over the entire Brillouin zone, see App.~\ref{sec:numerics}.
We obtained the same behavior as in the patch model.

\subsection{Ladder series}

The divergences that we found in $\chi_{\mathrm{pp}}^0$ and $\chi_{\mathrm{ph}}^0$ can lead to a pairing or to a $q=0$ instability in either spin or charge channel, when we separately sum up the corresponding ladder series.
We follow a standard protocol and introduce three types of infinitesimally small trial (bare) vertices $\Gamma^0_\mathrm{sc}$, $\Gamma^0_{\mathrm{s}}$, and $\Gamma^0_\mathrm{c}$, where $\mathrm{sc}$ stands for superconducting, and $\mathrm{s(c)}$ for spin (charge).
Because there are three non-equivalent HOVH points, each vertex is a three-component vector: ${\hat \Gamma}^0_i = (\Gamma^0_i (M_1), \Gamma^0_i (M_2), \Gamma^0_i (M_3))$ ($i =\mathrm{sc, c, s}$).

\begin{figure}[t!]
\begin{center}
\includegraphics[width=\columnwidth]{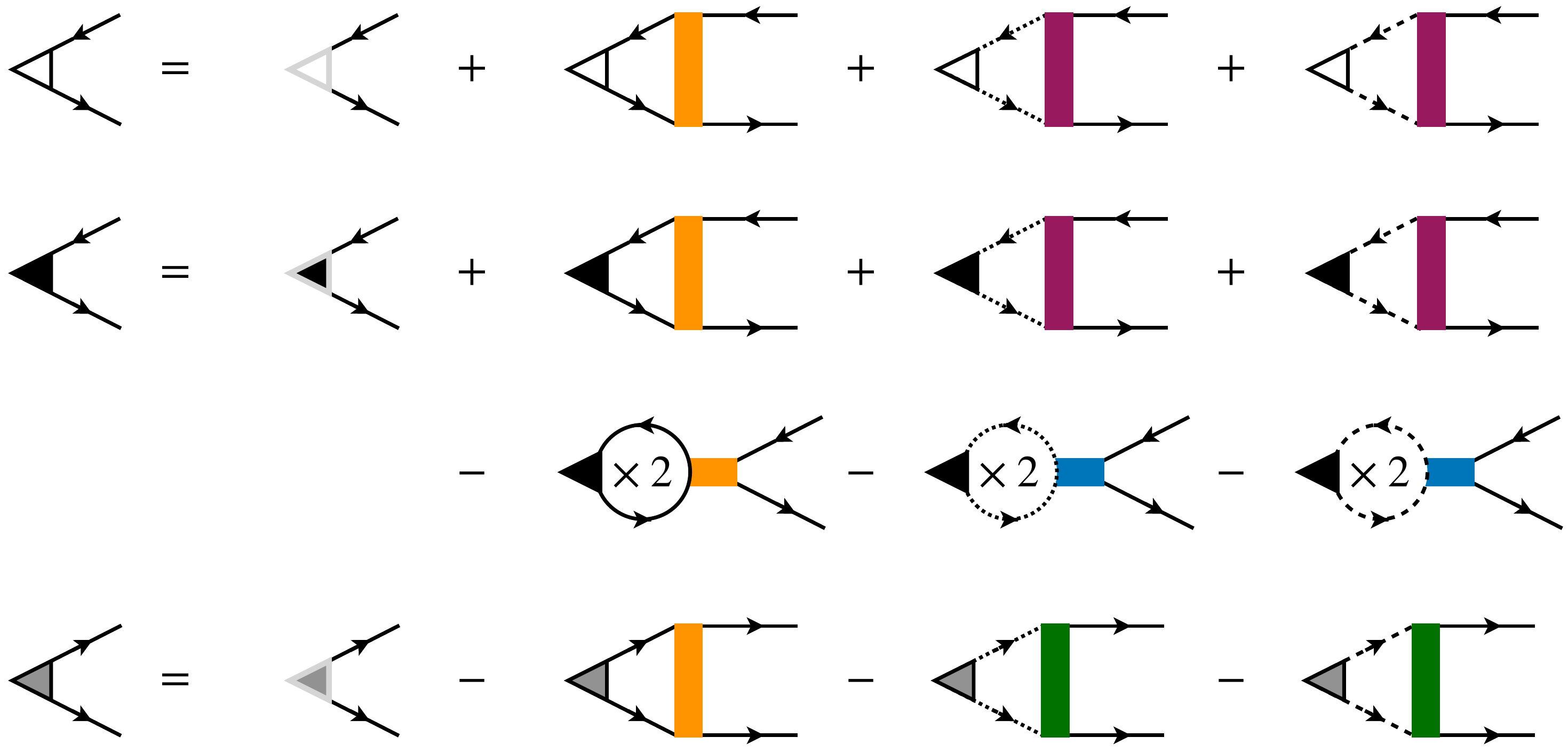}
\end{center}
\caption{\textbf{Ladder series for vertices.} Graphic representation of the ladder series for the spin (\textit{top row}), charge (\textit{middle rows}) and pairing (\textit{bottom row}) vertex. States close to the three $M_p$ points are respresented by solid, dashed and dotted lines. The couplings are colored according to Fig.~\ref{fig:gology}.}
\label{fig:vertexloops}
\end{figure}

The full vertices ${\hat \Gamma}_i$ are obtained by summing up ladder series of renormalizations.
In each section of a ladder we have the product of some combination of the couplings $g_i$ and either $\chi^0_{\mathrm{pp}}$ or $\chi^0_{\mathrm{ph}}$.
The ladder series are shown graphically in Fig.~\ref{fig:vertexloops}.
In analytical form, we obtain
\bea
   {\hat \Gamma}_{\mathrm{s}}  &=& {\hat \Gamma}^0_{\mathrm{s}} + {\hat \Gamma}_{\mathrm{s}} {\hat A}_{\mathrm{s}} \chi^0_{\mathrm{ph}}\,,\nonumber \\[3pt]
   {\hat \Gamma}_{\mathrm{c}}  &=& {\hat \Gamma}^0_{\mathrm{c}} + {\hat \Gamma}_{\mathrm{c}} {\hat A}_{\mathrm{c}} \chi^0_{\mathrm{ph}}\,,\\[3pt]
   {\hat \Gamma}_{\mathrm{sc}}  &=& {\hat \Gamma}^0_{\mathrm{sc}} + {\hat \Gamma}_{\mathrm{sc}} {\hat A}_{\mathrm{sc}} \chi^0_{\mathrm{pp}}\,,\nonumber
\label{ch_8}
\eea
where ${\hat A}_{i}$ are $3 \times 3$ matrices
\begin{align}
{\hat A}_{i}&=
\begin{pmatrix}
d_i & o_i & o_i \\
o_i & d_i & o_i \\
o_i & o_i & d_i
 \end{pmatrix}
\label{ch_9}
\end{align}
with the matrix elements being combinations of the couplings, i.e. $d_\mathrm{s}\!=\!g_4, o_\mathrm{s}\!=\!g_1$, $d_\mathrm{c}\!=\!-g_4, o_\mathrm{c}\!=\!g_1-2g_2$, $d_{\mathrm{sc}}\!=\!-g_4, o_{\mathrm{sc}}\!=\!-g_3$.
Each matrix equation can be decomposed into  three independent equations for the eigenvectors:
\bea
   \Gamma_{j,\mathrm{s}}  &=& \frac{\Gamma^0_{j,\mathrm{s}}}{1 - A_{j,\mathrm{s}} \chi^0_{\mathrm{ph}}}\,,\nonumber\\
   \Gamma_{j,\mathrm{c}}  &=& \frac{\Gamma^0_{j,\mathrm{c}}}{1 - A_{j,\mathrm{c}} \chi^0_{\mathrm{ph}}}\,,
   \label{ch_10}\\
    \Gamma_{j,\mathrm{sc}}  &=& \frac{\Gamma^0_{j,\mathrm{sc}}}{1 - A_{j,\mathrm{sc}} \chi^0_{\mathrm{pp}}}\,,\nonumber
\eea
where $j =1\ldots3$. We find
\begin{align}
A_{1,\mathrm{s}} &= g_4+2g_1,\ A_{2,\mathrm{s}}= A_{3,\mathrm{s}}= g_4 -g_1\,,\nonumber\\[3pt]
A_{1,\mathrm{c}} &= -g_4+2g_1-4g_2,\nonumber\\
A_{2,\mathrm{c}} &= A_{3,\mathrm{c}} = -g_4-g_1 +2g_2\,,\nonumber\\[3pt]
A_{1,\mathrm{sc}} &= -g_4-2g_3,\ A_{2,\mathrm{sc}}= A_{3,\mathrm{sc}} = -g_4 +g_3\,.
\end{align}
We see that the Fermi liquid state becomes unstable when $A_{j,i}\chi^0_{\mathrm{pp(ph)}}= 1$, i.e. at $T \sim |A_{j,i}|^{1/\kappa}$.
As $\chi^0_{\mathrm{pp}}$ and $\chi^0_{\mathrm{ph}}$ are of the same order, the type of the leading instability, i.e. whether it is superconducting or Pomeranchuk-type, and for which $j$, depends on the bare values of the couplings $g_1\ldots g_4$.

In the ladder approach, we consider each channel independently. This is the legitimate approximation if $A\chi^0$ in one particular channel is much larger than in other channels.
However, in our case, the susceptibilities in the particle-particle and the particle-hole channel are of the same order.
In this situation, the diagrams that couple different channels are of the same order as the ladder diagrams, and cannot be neglected.
Then we have to account for the mutual influence of fluctuations in different channels to correctly describe the low-energy behavior.

\section{Renormalization group}

To include the mixing between different channels, we employ a renormalization group (RG) approach, in which we keep all leading divergences at each loop order.
More formally, in a perturbation expansion the leading diagrams on the $n$-loop level will be proportional to $T^{n\kappa}$.
This includes $n$-loop diagrams from the different particle-particle and particle-hole ladders, but also mixed diagrams with insertions of a singular $l$-loop particle-particle contribution into a singular $(n-l)$-loop particle-hole diagram and vice versa ($l<n$).
The RG procedure approximates these mixed contributions by the product of decoupled $l$-loop particle-particle and $(n-l)$-loop particle-hole diagrams (or vice versa).
The analogous approximation appears for mixed diagrams of crossed and direct particle-hole type.
While this reproduces the correct temperature dependence, or, more generally, the dependence on the RG~scale, it introduces an inaccuracy in the prefactor of the mixed diagrams as typical moments in both channels are comparable, and the decoupling is justified only for the order-of-magnitude analysis.
The error is  formally controlled by the exponent $\kappa$ in the sense that for the logarithmic RG for $\kappa\rightarrow 0$, the decoupling is justified, to logarithmic accuracy.
To estimate the error introduced by the decoupling, we compute the two-loop mixed diagrams and compare them with the RG result in App.~\ref{sec:2loopdiagram}.
We find that the two are reasonably close to each other.
We therefore believe that the renormalization group approach, albeit approximate for HOVH points, is qualitatively accurate.

\subsection{RG equations}
\label{sec:RGeqs}

When setting up the RG procedure, it is important to choose a suitable regularization.
As we have shown in the previous section, the leading contributions come from bubbles with zero momentum transfer.
It is known that momentum-shell cutoffs can be disadvantageous for processes that involve small-momentum particle-hole fluctuations around the Fermi surface because they suppress these fluctuations by construction\cite{PhysRevB.64.184516,PhysRevB.79.195125}.
In a random phase approximation (RPA) treatment, this does not lead to problems, but in the description of the interplay of different ordering tendencies, particle-hole fluctuations with small and large momentum are not treated equivalently.
While this does not affect the competition of superconductivity and spin-density waves with large typical momentum, it is important in our case, where superconducting tendencies compete with zero-momentum orders.
Therefore, we choose an RG scheme in which the temperature regularizes interaction corrections and can be used as flow parameter\cite{PhysRevB.64.184516}.
Alternatively, one can use a frequency regularization scheme and integrate out modes with frequencies larger than a cutoff\cite{PhysRevB.79.195125,PhysRevB.85.035414}.
Eventually, both approaches yield the same renormalization group flow equations.

To systematically derive the RG equations, we start from a more general point of view and write down all possible vertex corrections within the patch model.
This does not only include the leading processes with characteristic momentum of zero, but also the subleading ones with momentum transfer $M_i$.
The flow equations read
\begin{align}
	\dot{g}_1=&-2\dot{\chi}^M_{\mathrm{pp}} g_1g_2+\dot{\chi}^0_{\mathrm{ph}}\left((N-2)g_1^2+2g_1g_4\right)\nonumber\\
	&-2\dot{\chi}^M_{\mathrm{ph}}g_1(g_1-g_2)\,,\label{betag1start}\\[5pt]
	\dot{g}_2=&-\dot{\chi}^M_{\mathrm{pp}}(g_1^2+g_2^2)+\dot{\chi}^M_{\mathrm{ph}}(g_2^2+g_3^2)\nonumber\\
	&-\dot{\chi}^0_{\mathrm{ph}}\left(2g_4(g_2-g_1)+2(N-2)g_2(g_2-g_1)\right)\,,\\[5pt]
	\dot{g}_3=&2\dot{\chi}^M_{\mathrm{ph}} g_3(2g_2-g_1)-\dot{\chi}^0_{\mathrm{pp}}\left(2g_3g_4+(N-2)g_3^2\right)\,,\\[5pt]
	\dot{g}_4=&-\dot{\chi}^0_{\mathrm{ph}}\left(2(N-1)g_2(g_2-g_1)-(N-1)g_1^2-g_4^2\right)\nonumber\\
	&-\dot{\chi}^0_{\mathrm{pp}}\left(g_4^2+(N-1)g_3^2\right)\,,\label{betag4start}
\end{align}
where the dots denote the derivatives with respect to the logarithm of the temperature $t=\ln \Lambda/T$, i.e. $\dot{g}_i=\frac{d}{dt} g_i$ and $\dot{\chi}_i^X=\frac{d}{dt} \chi_i^X$, and $\chi_i^{M}=\chi_i^{M_1}=\chi_i^{M_2}=\chi_i^{M_3}$ due to rotational symmetry.
In our case, the number of patches is $N=3$, but we keep $N$ as a parameter because the same set of RG equations holds for other cases, e.g., for the square lattice, where $N=2$.
We note in passing that these three-patch RG equations can be systematically derived from the more general functional RG equations, by restricting the possible scattering wave-vectors accordingly, see App.~\ref{sec:rg}.
Eqs.~\eqref{betag1start}-\eqref{betag4start} with $\kappa \to 0$ reproduce the logarithmic equations for a CVH points, cf.~Refs.~\onlinecite{PhysRevLett.81.3195,nandkishore2012chiral}.

\begin{figure}[t!]
\begin{center}
\includegraphics[width=\columnwidth]{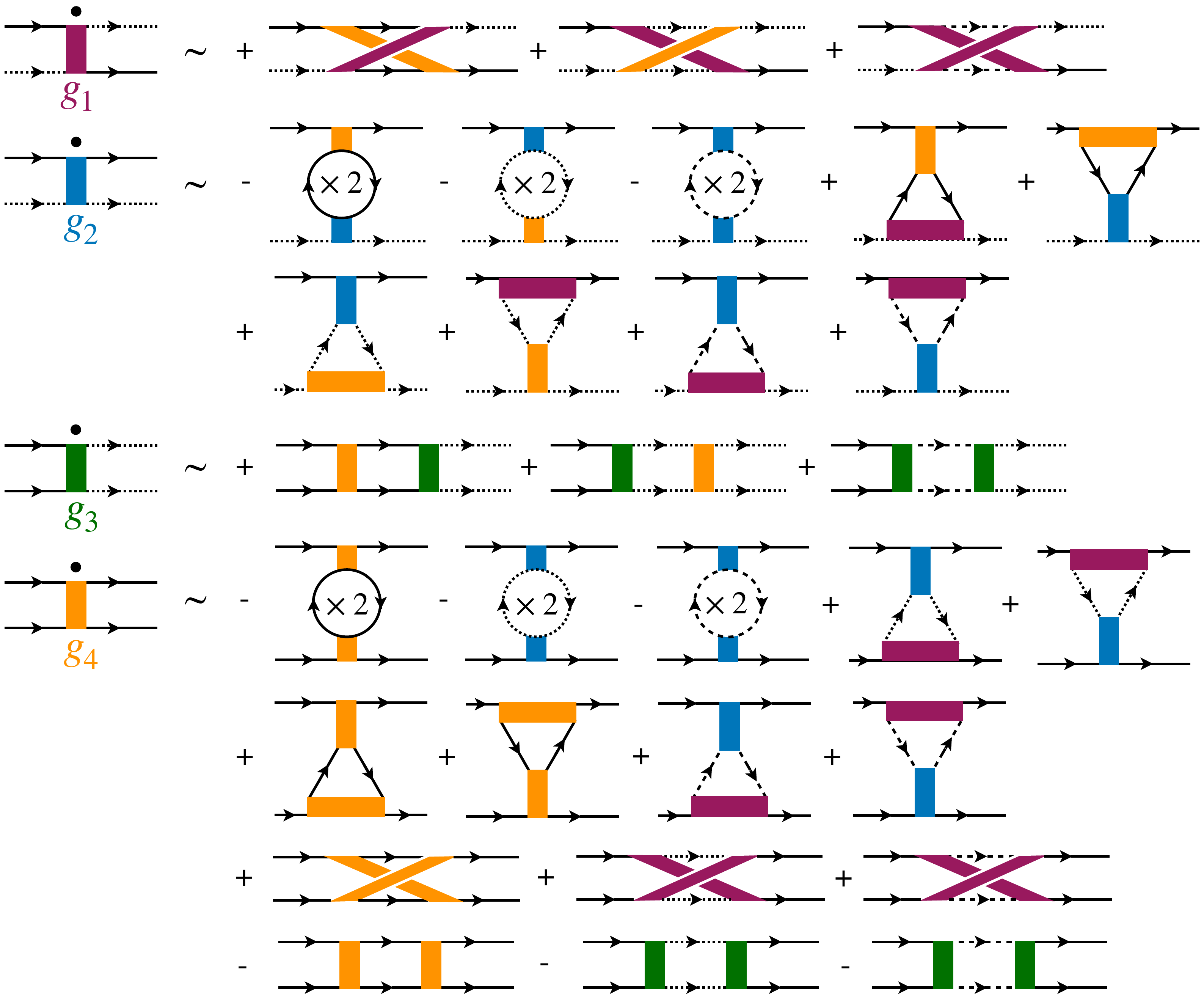}
\end{center}
\caption{\textbf{RG flow equations.} Diagrams representing the different RG flow Eqs.~\eqref{betag1b} -- \eqref{betag4b}. Note that the two internal lines correspond to the same $M$ point in each diagram. We show the flow equations for three $M_p$ points represented by solid, dashed and dotted lines.}
\label{fig:diagrams02}
\end{figure}

For $\kappa >0$, the leading terms in these equations are proportional to $\dot \chi_{\mathrm{pp}}^0$ and $\dot \chi_{\mathrm{ph}}^0$, which both scale as $T^{-\kappa}$.
We express their ratio as
\begin{align}
d_0=\chi^0_{\mathrm{ph}}/\chi^0_{\mathrm{pp}}\,.
\end{align}
Eqs.~\eqref{eq:chiph0} and \eqref{eq:chipp0} yield $d_0\approx 0.25$.
Below, we will use $d_0$ as a free parameter
to keep the equations applicable to other systems with HOVH points.
As we said, we neglect subleading
terms proportional to $\chi_i^M$ in Eqs.~\eqref{betag1start}-\eqref{betag4start}. We have checked numerically that the inclusion of constant $\dot \chi_i^{M}$, i.e. logarithmic $\chi_i^{M}$, does not change the results qualitatively.

\subsection{Dimensionless couplings}
\label{sec:dimless}

Keeping only $\dot \chi_{\mathrm{pp}}^0$ and $\dot \chi_{\mathrm{ph}}^0$ in (\ref{betag1start})-(\ref{betag4start}) and
introducing the dimensionless couplings $\hat g_i= g_i \partial_t \chi_{\mathrm{pp}}^0$, we obtain the flow equations for the case of $N$  HOVH points
\begin{align}
\partial_t \hat g_1&=\kappa\hat g_1+ d_0 \left[ (N-2)\hat g_1^2+2\hat g_1 \hat g_4\right]\,,\label{betag1t}\\
\partial_t \hat g_2&=\kappa\hat g_2+2d_0( \hat g_1- \hat g_2)\left[\hat g_4+ (N-2)\hat g_2\right]\,,\\
\partial_t \hat g_3&=\kappa\hat g_3- \hat g_3\left[2 \hat g_4+ (N-2)\hat g_3\right]\,,\\
\partial_t \hat g_4&=\kappa\hat g_4+d_0\left[ 2(N-1) \hat g_2(\hat  g_1- \hat g_2)+(N-1) \hat g_1^2+ \hat g_4^2\right]\nonumber \\
&\quad-( \hat g_4^2+(N-1) \hat g_3^2)\,.\label{betag4t}
\end{align}
We show a diagrammatic representation in Fig.~\ref{fig:diagrams02}.

For the case of a single HOVH point, the only available coupling is $g_4$. Setting $g_1=g_2=g_3=0$ and $N=1$ in Eq.~(\ref{betag4t}), we reproduce the RG equation in Ref. \onlinecite{PhysRevResearch.1.033206}:
$\partial_t \hat g_4=\kappa\hat g_4 - (1-d_0) \hat g_4^2$.
As demonstrated in Ref.~\onlinecite{PhysRevResearch.1.033206}, this equation has a non-trivial fixed point $\hat g_4^*=\kappa/(1-d_0)$, to which the system flows if the bare $\hat g_4$ is small enough (see also Ref.~\onlinecite{PhysRevB.95.035137}).
This fixed point describes a critical, metallic ground state -- the supermetal -- featuring power-law divergent charge and spin susceptibilities, but no long-range spin or charge order.
For more than one HOVH point, we find that the supermetal fixed point becomes unstable.
More generally, we searched for fixed points of Eqs.~\eqref{betag1t} --~\eqref{betag4t}, i.e. solutions with finite $\hat g_i$.
We find that all fixed points have at least one relevant direction in coupling space, i.e. they are all unstable.
The details of the calculation can be found in App.~\ref{sec:fps}.
We will search for fixed trajectories, instead, along which some couplings tend to infinity, indicating an instability of the ordinary metallic state.

\subsection{Flow to strong coupling}
\label{sec:strongflow}

\begin{figure}[t!]
\begin{center}
\includegraphics[width=\columnwidth]{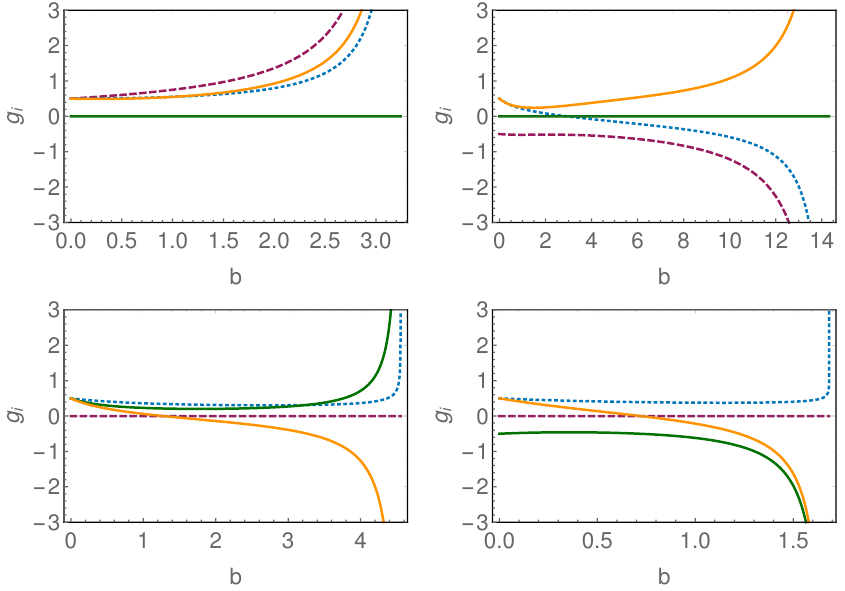}
\end{center}
\caption{\textbf{Flow to strong coupling.} Integration of the flow Eqs.~\eqref{betag1b} -- \eqref{betag4b} for four sets of bare couplings and $d_0=0.25$. The bare values are $g_1^0=g_2^0=g_4^0=0.5$, $g_3^0=0$ (\textit{top left}), $-g_1^0=g_2^0=g_4^0=0.5$, $g_3^0=0$ (\textit{top right}),  $g_2^0=g_3^0=g_4^0=0.5$, $g_1^0=0$ (\textit{bottom left}) and  $g_2^0=-g_3^0=g_4^0=0.5$, $g_1^0=0$ (\textit{bottom right}). Color code: $g_1$ (dashed, purple), $g_2$ (dotted, blue) , $g_3$ (solid, green), $g_4$ (solid, orange).}
\label{fig:flow}
\end{figure}

\begin{figure}[t!]
\begin{center}
\includegraphics[width=\columnwidth]{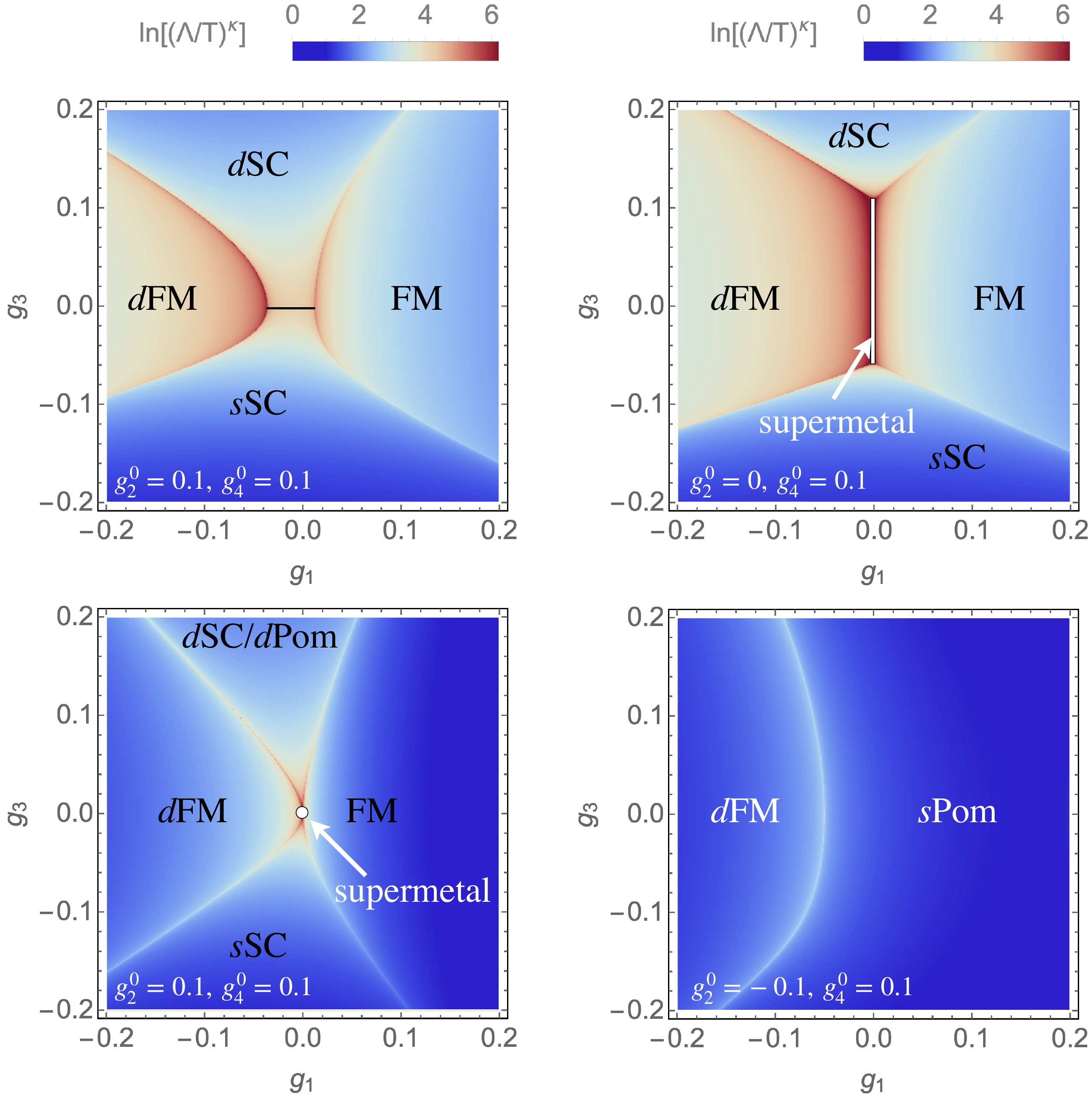}
\end{center}
\caption{\textbf{Phase diagrams.} Bare interactions $g_2^0=g_4^0$ are held fixed and $g_1$ and $g_3$ are varied,  $d_0=0.25$ (top) and $d_0=1$ (bottom). Bare values are $g_2^0=g_4^0=0.1$ (left) or $g_2^0=0$, $g_4^0=0.1$ (right, top), and $g_2^0=-0.1$, $g_4^0=0.1$ (right, bottom). The coloring encodes the scale where the couplings diverge, i.e. where correlations grow strong. If it is too large, e.g., in the red regime, there is no instability or it only occurs at the lowest scales. This can be used to estimate phase boundaries. We abbreviated FM: ferromagnet, $d$FM: $d$-wave spin Pomeranchuk, $s$POM: $s$-wave charge Pomeranchuk, $d$POM: $d$-wave charge Pomeranchuk, $s$SC: $s$-wave superconductivity, $d$SC: $d$-wave superconductivity. 
}
\label{fig:pdsketch}
\end{figure}

In the following, we determine the possible ground states of the system with more than one HOVH point.
We focus on our model with $N =3$.  As a convenient reparameterization we use as flow parameter 
\bea
b &=& 
\chi^0_{\mathrm{pp}} (T) - \chi^0_{\mathrm{pp}} (\Lambda)
= 
\frac{\rho_0 g(\kappa)}{\Lambda^\kappa }\left(\left(\frac{\Lambda}{T}\right)^\kappa -1\right)\,.
\label{new_n1}
\eea
We subtracted from $\chi^0_{\mathrm{pp}} (T)$ its value at the UV cutoff $\Lambda$ so that $b$ ranges from zero at the UV cutoff to infinity in the IR limit.
Using this $b$ as the RG scale and returning back to dimension-full couplings $g_i$, we obtain the compact flow equations
\begin{align}
\partial_b g_1&=d_0 ( g_1^2+2g_1 g_4)\,,\label{betag1b}\\
\partial_b g_2&=2d_0( g_1- g_2)(g_4+ g_2)\,,\\
\partial_b g_3&=- g_3(2 g_4+ g_3)\,,\\
\partial_b g_4&=d_0\left[ 4 g_2( g_1- g_2)+2 g_1^2+ g_4^2\right]-( g_4^2+2 g_3^2)\,.\label{betag4b}
\end{align}
The solution of this equation is shown  graphically in Fig.~\ref{fig:flow}.
We see that the running couplings diverge at a critical scale $b_c$, which signals an instability towards an ordered ground state.
Below, we discuss which instability develops first.
To reach the supermetal state, we have to fine-tune the bare values.
For example, we can set bare  $g_1 = g_2 =0$ and keep the bare $g_3$ within certain limits, see Fig.~\ref{fig:pdsketch}.
In this case, the couplings $g_1$ and $g_2$ remain zero, and $g_3$ and $g_4$ flow to zero as $T^\kappa$.
This means that the corresponding, rescaled dimensionless coupling $\hat g_{3,4} = \kappa g_{3,4} \chi^0_{\mathrm{pp}}$ approach fixed-point values.
Once the bare $g_{1}$ and/or $g_2$ are finite, the flow of the couplings is as in Fig. ~\ref{fig:flow}.

Note that we did not include the self-energy corrections into our RG equations.
The reason is that the contributions from the self-energy are subleading in their temperature dependence because the first non-analytic contribution to the self-energy $ \Sigma (T) \propto T^{1-2\kappa}$ appears at the two-loop order.
One can check that including such a self-energy into the diagrams for the renormalization of $g_i$ will only give rise to subleading terms.
Still, self-energy corrections can be relevant because they renormalize the chemical potential and can be expected to generate some additional quadratic momentum dependence in both directions of deviations from the Van Hove points.
Both effects spoil the HOVH behavior.
We absorb the renormalization of the chemical potential into the effective $\mu$, which we tune to the HOVH point.
We also assume that the scale $b_c$, at which the couplings diverge, is smaller than the one at which the momentum dependence, induced by the self-energy, becomes relevant.

\subsection{Fixed trajectories}

\begin{table*}[t!]
  \caption{\textbf{Fixed trajectories.} For convenience, we introduced the abbreviations $D_1=\sqrt{d_0(12+13d_0)}$, $D_3=\sqrt{9+8d_0}$, $D_5=\sqrt{d_0(13d_0-4)}$, and $D_7=\sqrt{d_0(-24d_0^2+85d_0-36)}$. See Fig.~\ref{fig:pdsketch} for abbreviations of instabilities. }
  \label{tab:FTs}
  \begin{tabular*}{\linewidth}{@{\extracolsep\fill}ccccccc}
    \hline\hline
    FT   & range of stability & $G_1$ &    $G_2$     &   $G_3$ & $G_4$& instability   \\
    \hline
    	(I)	 &     no restriction     &   $ \frac{-1+D_1}{d_0(13d_0-1)} $   &   $ G_1/2$      &  0  &   $\frac{13d_0-D_1}{2d_0(13d_0-1)}$& FM \\
    		(II)	 &     $d_0>1/13\approx 0.077$      &   $ \frac{-1-D_1}{d_0(13d_0-1)} $   &   $ G_1/2 $     &   $0$  &   $\frac{13d_0+D_1}{2d_0(13d_0-1)}$& d-FM \\
    	(III)	 &     $d_0<\frac{1}{16}(19+\sqrt{73})\approx 1.73$      &    0    &    0      &   $\frac{d_0-D_3}{9-d_0}$  &   $-\frac{4}{9+D_3}$ & s-SC\\
	(IV)	 &     $d_0<\frac{1}{16}(19-\sqrt{73})\approx 0.65$    &    0    &     0     &  $\frac{d_0+D_3}{9-d_0}$   &  $-\frac{4}{9-D_3}$ &d-SC \\
	(V)	 &     $d_0>\frac{1}{6}(9-\sqrt{33})\approx 0.54$     &    0    &     $\frac{-1+2d_0-D_5}{2d_0(1+3d_0)}$    &   0  &  $\frac{-5d_0+D_5}{2+6d_0}$  & s-POM\\
	(VI)	 &     $d_0>\frac{1}{6}(9+\sqrt{33})\approx 2.46$     &   0     &   $ \frac{-1+2d_0+D_5}{2d_0(1+3d_0)} $    &   0  &  $\frac{-5d_0-D_5}{2+6d_0}$  &d-POM\\
	(VII)	 &     $\substack{0.49\lesssim d_0\lesssim0.54 \\ 1.72\lesssim d_0\lesssim 3.05}$     &     0   &   $\frac{-9+10d_0-D_7}{6d_0(3+d_0)}  $    & $\frac{4d_0-3d_0^2-D_7}{3d_0(3+d_0)}$    &  $\frac{-13d_0+D_7}{6d_0(3+d_0)}$ &$\substack{\text{s-POM}\\ \text{s-SC/d-POM}}$\\
	(VIII)	 &     $0.65\lesssim d_0\lesssim 2.45$     &    0    &     $\frac{-9+10d_0+D_7}{6d_0(3+d_0)}  $    & $\frac{4d_0-3d_0^2+D_7}{3d_0(3+d_0)}$    &  $\frac{-13d_0-D_7}{6d_0(3+d_0)}$ &d-SC/d-POM\\
        \hline\hline
  \end{tabular*}
\end{table*}

When the couplings run into a singularity, they do so in a specific way, where the ratios of the couplings tend to finite values.
This is called a fixed trajectory (FT).
In general, there are several stable FTs and it depends on the bare couplings which one the system approaches.
Along a FT, the solutions of the RG equations follow the behavior
\begin{align}
\label{eq:geff}
g_i= G_i/(b_c-b)\,.
\end{align}
Solving the algebraic equations for $G_i$, we find different FTs.
We are interested in \textit{stable} FTs, to which the system flows under the RG for a range of bare couplings, i.e., without fine-tuning.
We find eight such stable trajectories for general $d_0$, see Tab.~\ref{tab:FTs}.
For $d_0\approx 0.25$, we can reach FTs (I) -- (IV).
We show the flow to these FTs in Fig.~\ref{fig:flow}.
When the UV cutoff is such that  $d_0\sim 1$, i.e. the susceptibilities in particle-hole and particle channels are about the same, FTs (V) and (VIII) also become available, while FT (IV) becomes unstable.

\subsection{RG-enhanced susceptibilities}

Next, we use the information about the fixed trajectories to study how the susceptibilities for different ordering tendencies behave.
To that end, we again introduce the trial vertices $\hat\Gamma_{i}$ in superconducting, and spin and charge $q=0$ channels ($i\in\{sc,s,c\}$).
We rewrite Eq.~\eqref{ch_8} as a differential equation, i.e.
\begin{align}
\partial_b\hat\Gamma_{i}&= \hat A_i  \hat\Gamma_{i}  \,,
\label{new_n2}
\end{align}
where  ${\hat A}_{i}$ are the $3 \times 3$ matrices still given by Eq.~(\ref{ch_9}), but the couplings now are the running ones -- the solutions of the RG Eqs.~\eqref{betag1b} -- \eqref{betag4b}, which include the contributions from mixed diagrams.

Solving Eq.~\eqref{new_n2}, we find that the eigenvectors $\Gamma_{j,i}$, $j\in\{1,2,3\}$ diverge as $(b_c-b)^{-\beta_{j,i}}$, where the exponents $\beta_{j,i}$ are expressed via the parameters $G_1,G_2,G_3,G_4$  characterizing the fixed trajectories.
Out of the three $\Gamma_{j,i}$ for each $i$, $\Gamma_{1,i}\propto (1,1,1)$  corresponds to $s$-wave symmetry, and $\Gamma_{2,i}\propto(0,1,-1)$ and $\Gamma_{3,i}\propto(1,-1/2,-1/2)$ correspond to $d$-wave symmetry.
For the latter, the exponents are degenerate~\cite{nandkishore2012chiral,maiti2013superconductivity}.
We label the exponents as $\beta_{1,i}=\beta_i^{(s)}$ and $\beta_{2,i}=\beta_{3,i}=\beta_i^{(d)}$ and find
\begin{align}
\beta_{\mathrm{s}}^{(s)}&=d_0(G_4+2G_1)\,,\\
\beta_{\mathrm{s}}^{(d)}&=d_0(G_4-G_1)\,,\label{exps}\\
\beta_{\mathrm{c}}^{(s)}&=d_0(-G_4+2G_1-4G_2)\,,\\
\beta_{\mathrm{c}}^{(d)}&=d_0(-G_4-G_1+2G_2)\,,\\
\beta_{\mathrm{sc}}^{(s)}&=-G_4-2G_3\,,\\
\beta_{\mathrm{sc}}^{(d)}&=-G_4+G_3\label{expsc}\,.
\end{align}
The corresponding susceptibilities behave like
\begin{align}
\chi_{j,i}\propto\int db \,  \Gamma_{j,i}^2\propto (b_c-b)^{1-2\beta_{j,i}}\,,
\end{align}
see Refs.~\onlinecite{Metzner,Vafek_14,Chubukov_16,Classen_17_1,Classen_17_2} for earlier discussions on this issue.
The leading instability at $b = b_c$  will be into the ordered state  for which $\beta_{j,i}$ is the largest.
Comparing the exponents on different fixed trajectories, cf. App.~\ref{sec:exponents},
we find that the following orders develop, depending on the bare couplings:
\begin{enumerate}
\item[(I)] ferromagnetism
\item[(II)] $d$-wave spin Pomeranchuk order
\item[(III)] $s$-wave superconductivity
\item[(IV)] $d$-wave superconductivity
\item[(V)] $s$-wave charge Pomeranchuk order
\item[(VI)] $d$-wave charge Pomeranchuk order
\item[(VII)]
for $0.49\!\lesssim\!d_0\!\lesssim\!0.54$ $s$-wave charge Pomeranchuk,\\
for $1.72\!\lesssim\!d_0\!\lesssim\! 2.41$ $s$-wave superconductivity,\\
for $2.41\lesssim 3.05$ $d$-wave charge Pomeranchuk order
\item[(VIII)]
for $0.65\lesssim d_0<1$, $d$-wave superconductivity,\\
for $1\!<\!d_0\!\lesssim\!2.45$, $d$-wave charge Pomeranchuk order
\end{enumerate}
Based on this analysis, we can now determine the phase diagram by solving the flow equations for various bare couplings.
The result is shown in Fig.~\ref{fig:pdsketch}.
We consider different ranges of the bare couplings to map out all possible instabilities.
For a specific lattice model with onsite interaction $U$, nearest-neighbor interaction $V$ and  nearest-neighbor spin exchange $J$ on the honeycomb lattice, we obtain $g_1^0=U-V/2-J$, $g_2^0=U+3V/2-J$, $g_3^0=U-V/2+J$ and $g_4^0=U+3V/2-3J$.
However, note that the bare values of the patch model can be altered from the microscopic interactions due to modes with energies higher than the UV cutoff.

We find that for  purely repulsive bare couplings, the leading instabilities for $d_0\sim 0.25$ are ferromagnetism and $d$-wave superconductivity.
Superconductivity is driven by the pair-hopping term $g_3$, which needs to be sufficiently larger than $g_1$.
For larger $d_0$, the ferromagnetic region grows.
For $d_0>1$, d-wave superconductivity is replaced by $d$-wave charge Pomeranchuk order.

In case some couplings become attractive, we find an $s$-wave pairing, charge Pomeranchuk order, and a tendency towards a $d$-wave spin Pomeranchuk order.
In fact, only a small negative $g_1$ is needed to induce the $d$-wave spin Pomeranchuk order.
We also find the supermetal phase~\cite{PhysRevResearch.1.033206}, where couplings do not diverge in our phase diagram.
However, as we explained in Sec.~\ref{sec:RGeqs}, the bare couplings must be tuned to certain values to reach this phase, see Fig.~\ref{fig:pdsketch}.

\section{Free energy for $d$-wave orders}

Within the RG analysis, the exponents in the $d$-wave channels $\beta_{2,i}$ and $\beta_{3,i}$ are equal.
Hence, the system simultaneously becomes unstable towards the order with the structure set by the (normalized)  $\Gamma_{2}=1/\sqrt{2} (0,1,-1)$ and $\Gamma_{3}=\sqrt{2/3} (1,-1/2,-1/2)$.
The corresponding order parameters are commonly called $d_{xy}$ and  $d_{x^2-y^2}$ due to their symmetry.
To determine which combination of the $d_{xy}$ and $d_{x^2-y^2}$ orders develops, one needs to analyze the Landau free energy.

For a SC order, we introduce $\Delta_{\mathrm{sc}}=\Delta_2 \Gamma_{2}+\Delta_3 \Gamma_{3}$.
Both $\Delta_1$ and $\Delta_2$ are  $U(1)$ complex order parameters. The free energy is of the form
\begin{align}
\mathcal{F}_{d\mathrm{SC}} =& \frac{\alpha}{2}\left(\left|\Delta_2\right|^2+\left|\Delta_3\right|^2\right) \nonumber \\
&+\beta_1\left(\left|\Delta_2\right|^2+\left|\Delta_3\right|^2\right)^2 + \beta_2\left| \Delta_2^2 + \Delta_3^2\right|^2\,.
\end{align}
The coefficient $\alpha$ changes sign at the transition. We verified that for a HOVH point, $\beta_1,\beta_2>0$, like for the case of a CVH point in graphene~\cite{nandkishore2012chiral}.
In this case, the combination $\Delta_3=\pm i\Delta_2$ minimizes the free energy, i.e. $\Delta_{\mathrm{sc}}=(\Gamma_2\pm i\Gamma_3)\Delta$ with complex $\Delta$.
This is a chiral $d\pm i d$ superconducting state~\cite{black2014chiral}.

For $d$-wave charge Pomeranchuck order, we introduce two real order parameters
$\varphi_2$, $\varphi_3$ and the total order parameter is $\varphi_\mathrm{c}=\varphi_2 \Gamma_2+\varphi_3\Gamma_3$.
A hexagonal lattice allows for a cubic term in the free energy~\cite{PhysRevLett.110.126405,hecker2018vestigial,little2019observation,jin2019dynamical,PhysRevB.98.245103,2019arXiv191111367F}.
Keeping this term and neglecting $\varphi^4$ terms, we obtain
\begin{align}
\mathcal{F}_{d\mathrm{POM}} = \frac{\bar\alpha}{2}\left(\varphi_2^2+\varphi_3^2\right) + \bar \beta\left(\varphi_3^3-3\varphi_3\varphi_2^2\right) \,.
\end{align}
Minimizing the free energy (including quartic terms), we find that the system chooses one out of three equivalent states:
either $\varphi_c\propto (2, -1,-1)$, $\varphi_c\propto(-1,2, -1)$, or  $\varphi_c\propto(-1,-1,2)$.
Each state selects one particular HOVH point where the order is largest. 
Such a state breaks lattice $C_3$ rotational symmetry and is a charge nematic~\cite{PhysRevLett.110.126405,hecker2018vestigial,little2019observation,jin2019dynamical,PhysRevB.98.245103,2019arXiv191111367F}.

\begin{figure}[t!]
\begin{center}
\includegraphics[width=\columnwidth]{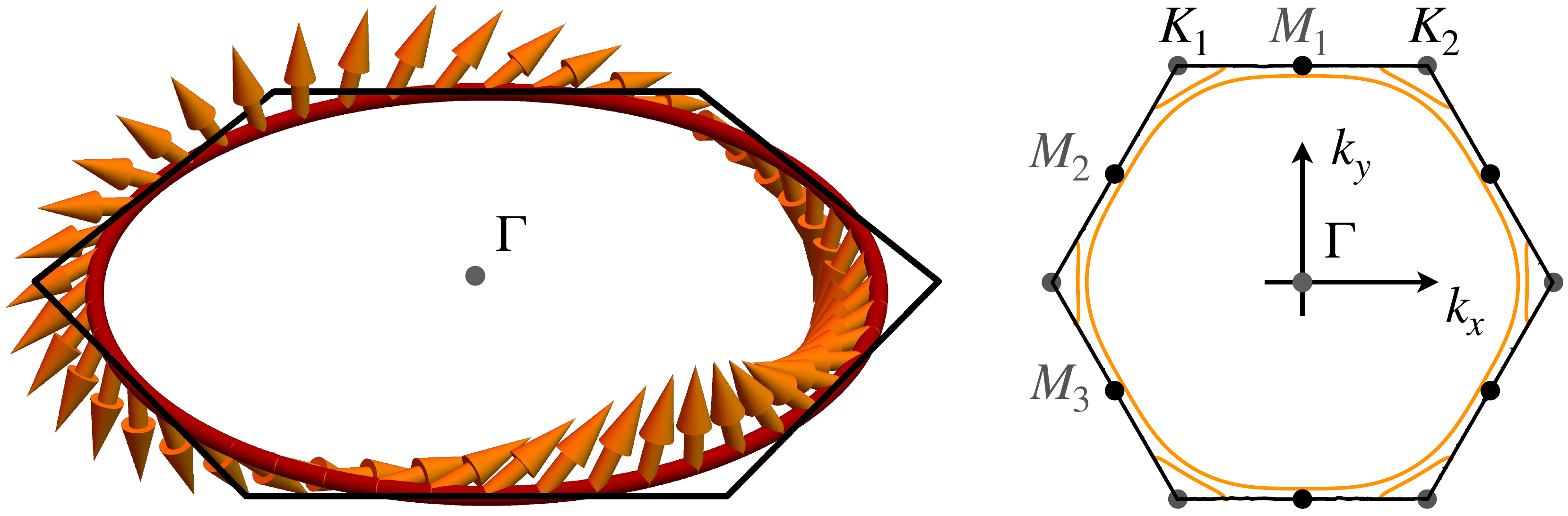}
\end{center}
\caption{\textbf{$d$-wave spin Pomeranchuk order.} \textit{Left panel:} the order parameter of the d-wave spin Pomeranchuk order winds twice around the Fermi surface. \textit{Right panel:} diagonalization leads to a Zeeman-like term without net magnetization which splits the Fermi surface.}
\label{fig:spinPomFS}
\end{figure}

For the $d$-wave spin channel, we express the order parameter via $O(3)$-symmetric vectors
$\vec \phi=\vec\phi_{2}\Gamma_2 +\vec \phi_3\Gamma_3$ ($\phi_i$ and $\Gamma_i$ live in different vector spaces).
The cubic term is absent and the free energy up to quartic order in $\vec \phi_{2,3}$ is given by the expression
\begin{align}
\mathcal{S}_{d\mathrm{FM}}&=\frac{\tilde \alpha}{2}\left(\vec{\phi}_1^2 +\vec{\phi}_2^2 \right ) \nonumber \\
&+ \frac{\tilde\beta}{2} \left[\left(\vec{\phi}_1^2\!+\vec{\phi}_2^2 \right)^2 - \frac{4}{3}\vec{\phi}_1^2\vec{\phi}_2^2 + \frac{4}{3} \left(\vec{\phi}_1\cdot\vec{\phi}_2 \right)^2 \right].
\end{align}
The coefficient $\tilde\beta$ is obtained by integrating out the fermions near a HOVH point. We find a positive result, reading
\begin{align}
\tilde\beta=\int G^4&=T\sum_\omega\int\,d\epsilon \frac{\rho(\epsilon)}{6}\partial_\epsilon^3 \frac{1}{i\omega-\epsilon}\\
&=\frac{\rho_0}{48}T^{-(2+\kappa)}\int\, du \frac{2-\cosh u}{|u|^\kappa\cosh^4(\frac{u}{2})}>0\,.
\end{align}
For $\kappa=1/4$, we have $\beta\approx 0.96/T^{9/4}$. Minimizing the free energy, we find that both $\vec\phi_2$ and $\vec\phi_3$ are non-zero.
Specifically, $|\vec\phi_2|=|\vec\phi_3| = |\vec\phi|$, and $\vec\phi_2$ and $\vec\phi_3$ are perpendicular to each other.
Combining this with the $d$-wave modulation, and extending the modulation to the full Fermi surface, we find that the spin order parameter winds twice around the Fermi surface.
We illustrate this in Fig.~\ref{fig:spinPomFS}.
This order breaks $SU(2)$ spin symmetry, but does not generate net magnetization due to the $d$-wave form factor.

To see the effect of the $d$-wave spin Pomeranchuk order on the energy dispersion, we consider the mean-field Hamiltonian
\beq
\label{eq:MF}
H_{\mathrm{MF}}=\sum_{p\sigma}\epsilon(\vec p) c_{p\sigma}^\dagger c_{p\sigma}+g_{s,d}\sum_{p\sigma,\sigma'}\vec\phi(\vec p) c^\dagger_{p\sigma}\vec\sigma_{\sigma\sigma'} c_{p\sigma'}\,,
\eeq
with free energy dispersion $\epsilon(\vec p)$ and coupling $g_{s,d}$.
We express the order parameter $\vec\phi(\vec p)$ as  $\vec\phi(\vec p)=d_{xy}(\vec p) \hat\phi_2+ d_{x^2-y^2}(\vec p)\hat \phi_3$, where $\hat\phi_2$ and $\hat\phi_3$ are orthogonal order parameters and $d_{xy}(\vec p)$, $d_{x^2-y^2}(\vec p)$ are the momentum-dependent $d$-wave form factors $d_{xy}=\sin2\theta_p$ and $d_{x^2-y^2}=\cos2\theta_p$ with the polar angle $\theta_p$.

$H_{\mathrm{MF}}$ is non-diagonal in the spin index. Diagonalization leads to the reconstructed energy bands
\beq
 E(\vec k)=\epsilon(\vec k)-\mu\pm|\vec\phi|\,.
\eeq
We see that the $d$-wave spin Pomeranchuk order introduces a Zeeman-like splitting, cf. Fig.~\ref{fig:spinPomFS}, which is, however, not spin-polarized.
In real space, the $d$-wave form factor translates to a modulation of the nearest-neighbor hopping.
As $\vec\phi$ couples to the electron spin, see Eq.~\eqref{eq:MF}, the hopping becomes spin-dependent.

\section{Conclusion}

We presented an analysis of competing instabilities for a system of interacting electrons in the presence of multiple HOVH points.
At a HOVH point, the density of states diverges by a power law, and we have shown that this gives rise to a qualitatively new type of competition between superconducting and zero-momentum particle-hole orders.
Our analysis of particle-particle and particle-hole susceptibilities has revealed that the ones with zero momentum transfer diverge by a power law, with the same exponent in the particle-particle and the particle-hole channel, while the ones at a finite momentum transfer diverge at most logarithmically.
This is in sharp contrast to CVH points, where the divergences are logarithmic, and the susceptibilities in the particle-hole channel are either subleading to the ones in the particle-particle channel, or are comparable, but at finite momentum transfer, if the Fermi surface is nested.

We argued that the physics associated with multiple HOVH points in the Brillouin zone, is relevant for intercalated graphene, where Van Hove filling has recently been achieved experimentally.
It was observed that near this filling, the band dispersion is strongly flattened around the Van Hove points\cite{PhysRevB.100.121407}.
HOVH points also appear in, e.g. twisted bilayer graphene, where they can be accessed by single-parameter tuning\cite{yuan2019magic}.

To model the HOVH scenario in graphene-based systems, we introduced a tight-binding model on the honeycomb lattice with up to third-nearest neighbor hopping and tuned the hopping amplitudes such that higher-order saddle points appear at the three inequivalent $M$ points in the Brillouin zone.
We derived the effective patch model for electrons around the HOVH points, which includes couplings for all symmetry-allowed scattering processes.
To analyze the competition between different ordering channels in an unbiased way, we set up the renormalization group approach that accounts for all leading fluctuation corrections.
The patch model and the renormalization group equations are valid for both hexagonal and tetragonal systems with HOVH points at the Brillouin zone edges.

We have shown that the supermetal state, which was predicted to be the ground state for a single HOVH point, is unstable when several HOVH points are present.
It can only survive under special fine-tuned conditions for the initial couplings.
For generic initial conditions, we observed a flow to strong coupling, indicating that an initial Fermi-liquid state becomes unstable towards a symmetry-broken ordered state.
We obtained the phase diagram for parameters relevant for intercalated graphene.
It includes  regions of ferromagnetism,
charge and
 spin Pomeranchuk orders, as well as $s$- and $d$-wave superconductivity.
The development of a specific instability depends on model parameters, i.e. the bare couplings and the ratio of the particle-particle and the particle-hole susceptibilities.
For purely repulsive interactions, we found that two key competitors are ferromagnetism and chiral $d+id$ superconductivity.  
We note that slightly away from Van Hove filling, spin-triplet $f$-wave superconductivity can also become a competitor\cite{nandkishore2014superconductivity,PhysRevB.86.020507,PhysRevB.99.201106,PhysRevB.99.195120}. We expect this tendency to be stronger in the vicinity of HOVH points because of increased ferromagnetic fluctuations. 
If some interactions turn attractive, $s$-wave superconductivity can develop.
In addition, we found that under some initial conditions the system develops $d$-wave charge or spin Pomeranchuk order.
We analyzed the free energy for the $d$-wave Pomeranchuk orders to determine the ground state configurations.
We found that the $d$-wave charge Pomeranchuk order breaks lattice rotational symmetry.
 For the $d$-wave spin Pomeranchuk order, we found that the order parameter winds twice around the Fermi surface.
Such an order is very unconventional:
it breaks spin SU(2) symmetry and splits the Fermi surface, but it does not introduce a net magnetization.
Our results demonstrate that the many-body phase diagram of
intercalated graphene and similar systems is very rich and hosts not only chiral superconductivity, but also unconventional spin and charge orders.

In future work, it will be interesting to improve our RG procedure regarding self-energy corrections or the approximation error of the mixed diagrams, by, e.g., employing functional RG techniques with more sophisticated truncations.
Another future research direction is to adapt our formalism to systems that possess HOVH points in different locations of the Brillouin zone.
One straightforward application is to the case of twisted bilayer graphene, where three HOVHs lie along the $\Gamma-M$ line away from the zone boundary~\cite{yuan2019magic}.

\subsection*{Acknowledgments}
%
We thank D.~V.~Chichinadze, D.~M.~Kennes, R.~Thomale, A.~M.~Tsvelik, and S.~Wessel for valuable discussions.
MMS was supported by the DFG through SFB 1238 (projects C02 and C03, project id 277146847) and CH through RTG 1995.
LC was supported by the Humboldt foundation during the first part and by the U.S. Department of Energy (DOE), Office of Basic Energy Sciences, under Contracts No. DE-SC0012704 during the later part of the project.
AVC was  supported by the Office of Basic Energy Sciences, U.S. Department of Energy, under award DE-SC0014402.

\begin{appendix}

\section{Band dispersion near the $M$ points}
\label{app:Mexp}

To demonstrate how the HOVH points come about in our model, we expand the dispersion around the $M$ points. For the expansion at $M_1=\pi(0,2/3)$, we obtain
\begin{align}
	\epsilon_{M_1}(\vec{x})= -a_1x^2+b_1y^2+c_1x^4+d_1y^4+e_1 x^2y^2+...\,,\nonumber
\end{align}
where $\vec{x}=(x,y)=(k_x-M_{1,x},k_y-M_{1,y})$ is the deviation from the corresponding $M$ point and the dots denote higher order terms in $x,y$. The coefficients are
\begin{align}
	a_1=&\frac{3}{4}\left(t_1-2t_2-4t_3\right)\,,\\
	b_1=&\frac{2 t_1^2}{t_1-3 t_3}+\frac{t_1}{4}+3 t_3-\frac{9 t_2}{2}\,,\\
	c_1=&\frac{3}{64} \left(t_1-2 \left(7 t_2+8 t_3\right)\right)\,,\\
	d_1=&\frac{3(9t_2-8t_3)}{32}+\frac{7 t_3 t_1^2-63 t_1^3+27 t_3^2 t_1+405 t_3^3}{64\left(t_1-3 t_3\right){}^3/(3t_1)}\,,\nonumber\\
	e_1=&\frac{27}{16} t_2+\frac{27 \left(t_1^3-14 t_3 t_1^2+33 t_3^2 t_1-16 t_3^3\right)}{32\left(t_1-3 t_3\right){}^2}\,.
\end{align}
The energy dispersion near the other $M$ points, $M_2=\pi(-1/\sqrt{3},1/3)$ and $M_3=-\pi(1/\sqrt{3},1/3)$, is
\begin{align}
	\epsilon_{M_2}(\vec{x})&= a_2 x^2+c_2 x y+ b_2 y^2+...\,,\\
	\epsilon_{M_3}(\vec{x})&= a_3 x^2+c_3 x y+ b_3 y^2+...\,,
\end{align}
where $\vec{x}$ is again measured from the corresponding $M$ point and the coefficients are
\begin{align}
	a_2&=a_3=\frac{3 t_1^2}{2 \left(t_1-3 t_3\right)}-3 t_2+3 t_3\,,\\
	b_2&=b_3=\frac{9 \left(t_1-2 t_3\right) t_3}{2 \left(t_1-3 t_3\right)}\,,\\
	c_2&=-c_3=-\frac{3 \sqrt{3} \left(t_1^2-\left(2 t_2+t_3\right) t_1+6 t_2 t_3\right)}{2 \left(t_1-3 t_3\right)}\,.
\end{align}
%

\subsection*{High-order saddle point}

We note that the quadratic term $\propto x^2$ in Eq.~\eqref{eq:m1approx} can be tuned to zero by choosing
\begin{align}
	t_3 \to t_{3,\mathrm{e}}=\left(t_1-2t_2\right)/4\,.
\end{align}
In that case, the usual saddle point is replaced by an even flatter energy dispersion. More explicitly, the band dispersion near $M_1$ then reads
\begin{align}
\epsilon_{M_1}(\vec{x})\big|_{t_{3,e}}= \bar b_1y^2-\bar c_1x^4-\bar e_1x^2y^2+...\,,
\end{align}
where we have introduced
\begin{align}
	\bar b_1&= t_1-6t_2+\frac{8t_1^2}{t_1+6t_2}\,,\\
	\bar c_1&=\frac{9}{64}(t_1+2t_2)\,,\\
	\bar e_1&=\frac{27(t_1+2t_2)\left(11 t_1^2 - 28 t_1 t_2 - 52 t_2^2\right)}{32(t_1+6t_2)^2}\,.
\end{align}
For these parameters we not only have $\nabla_{\vec{x}}\epsilon_{M_1}(\vec{x})=0$ at the saddle point, but also the Hessian matrix $H_{ij}=\partial_{x_i}\partial_{x_j} \epsilon_{M_1}(\vec{x})$ has a vanishing determinant, i.e. $\det H(\vec{x})=0$. The higher-order saddle point is shown in Fig.~\ref{fig:hosp}.
%

\subsection*{Density of states}

At such a two-dimensional higher-order saddle point, the DOS shows a power-law divergence
\begin{align}
	\rho(\epsilon)\propto |\epsilon|^{-\kappa}\,,
\end{align}
with some exponent $\kappa>0$. This divergence is stronger than the logarithmic one at a CVH singularity.
The singular behavior of the DOS near the high-order saddle point can be determined from a scaling argument~\cite{yuan2019classification}.
To that end, the Taylor expanded dispersion $\epsilon_{M_1}(\vec{x})$ is decomposed into two parts, the canonical part $\epsilon_{c}(\vec{x})$ and a perturbation $\epsilon_{p}(\vec{x})$, i.e. $\epsilon_{M_1}(\vec{x})=\epsilon_{c}(\vec{x})+\epsilon_{p}(\vec{x})$.
The canonical part has vanishing gradient and is defined by being scale invariant
\begin{align}
	\epsilon_{c}(\lambda^p x,\lambda^q y)=\lambda \epsilon_{c}(x,y)\,.
\end{align}
The perturbation part can, again, be decomposed into monomials with individual scaling behavior reading
\begin{align}
	\epsilon_{p}(\lambda^p x,\lambda^q y)=\lambda^r \epsilon_{p}(x,y)\,.
\end{align}
The perturbation is irrelevant at the HOVH point for $r>1$ and relevant for $r<1$. In our present scenario, we have
\begin{align}
	\epsilon_{c}(x,y)&=-\bar c_1x^4+\bar b_1y^2\, \Rightarrow\ p=1/4,\ q=1/2\,,\\
	\epsilon_{p}(x,y)&=-\bar e_1x^2 y^2+...\,,
\end{align}
so the scaling exponent of the monomial $\propto \bar e_1$ is $r=3/2 > 1$ and therefore it is irrelevant as well as all higher-order terms.
Using the canonical dispersion, the scale invariance and the definition of the DOS $\rho(\epsilon)=\int_{\vec{x}}\delta(\epsilon- \epsilon_c (\vec{x}))$, one can show that the DOS is also scale invariant
\begin{align}
	\rho(\lambda\epsilon) = \lambda^{\kappa}\rho(\epsilon)\,,
\end{align}
with $\kappa=p+q-1$ and $p, q$ as determined above. Then, the singular part of the DOS behaves according to
\begin{align}
	\rho(\epsilon)\propto |\epsilon|^{\kappa}\propto |\epsilon|^{-\frac{1}{4}}\,,
\end{align}
i.e. in our model $\kappa=1/4$. 

\section{Numerical evaluation of the loops}
\label{sec:numerics}

\begin{figure}[t]
\begin{center}
\includegraphics[width=\columnwidth]{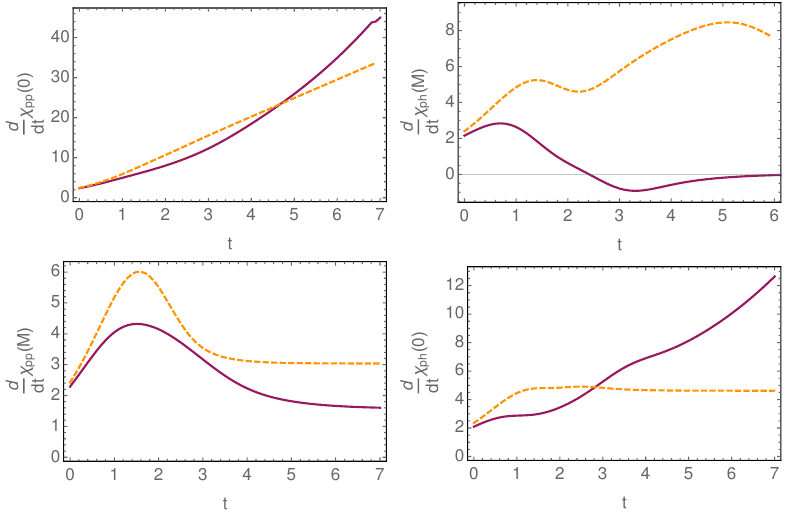}
\end{center}
\caption{\textbf{Temperature-derivative of loop corrections.}
We plot $d/dt \chi^X_{\mathrm{pp}}$ and $d/dt \chi^X_{\mathrm{ph}}$, cf. Eqs.~\eqref{eq:chipp}-\eqref{eq:chiph} with $t=\ln\Lambda/T$ for the microscopic model Eq.~\ref{eq:model}. \textit{From left to right and top to bottom}, we show $d/dt \chi^0_{\mathrm{pp}}$, $d/dt \chi^M_{\mathrm{ph}}$, $d/dt \chi^M_{\mathrm{pp}}$, and $d/dt \chi^0_{\mathrm{ph}}$. As an example for the situation with a HOVH singularity we chose $t_3=t_1/4, t_2=0$ (solid, purple). For comparison, we also show the case with a logarithmic DOS for $t3=0.01t_1, t_2=0$ (dashed, yellow). When plotted as function of the logarithmic temperature $t$, $\chi_i^X\propto (\Lambda/T)^\kappa$ leads to an exponential growth as observable in $\chi^0_{\mathrm{pp}}$ and $\chi_{\mathrm{ph}}^0$ (solid, purple). The logarithmic singularity $\chi_i^X\propto \ln^2\Lambda/T$ leads to a linear behavior in the derivative as observable for $\chi_{\mathrm{pp}}^0$ and $\chi_{\mathrm{ph}}^M$ (dashed, yellow). The linear growth in $\chi_{\mathrm{ph}}^M$ is cut for small enough temperatures $t\gtrsim 5$ due to imperfect nesting $t_3=0.01t_1$.}
\label{fig:dtbubbles}
\end{figure}

\begin{figure}[t!]
\begin{center}
\includegraphics[width=0.7\columnwidth]{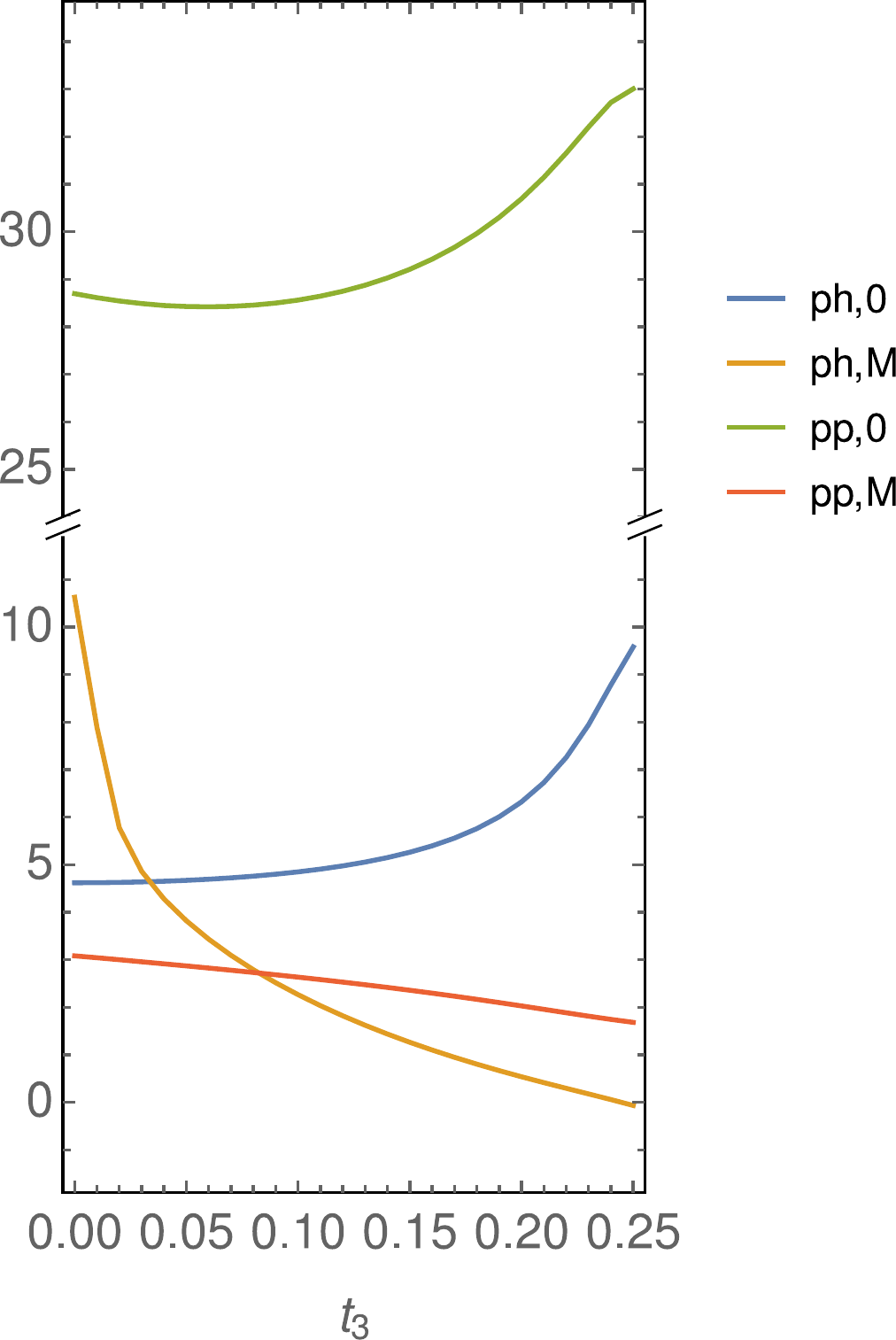}
\end{center}
\caption{\textbf{Loop derivatives} $\partial_t \chi^X_{\mathrm{pp/ph}}$ as function of $t_3$ for a fixed (arbitrary) temperature $t=\ln\Lambda/T=5.8$.}
\label{fig:dtBt3}
\end{figure}

We confirm the hierarchy of particle-particle and particle-hole bubbles that we obtained through expansion around the Van Hove points by numerical evaluation of the loops in the full Brillouin zone with the microscopic dispersion.
We use the third-neighbor hopping as the tuning parameter to change from a logarithmic divergence to a power-law divergence at the van-Hove points.
Fig.~\ref{fig:dtbubbles} shows the temperature-dependence of $\partial_t \chi^X_{\mathrm{pp/ph}}$ (Eq.~\eqref{eq:chipp}) for two cases:
an almost perfectly nested situation with $t_3=0.01t_1$ and the situation with an eVHS for $t_3=t_1/4$ ($t_2=0$ in both cases).
Because we plot the loop derivatives against the logarithm of the temperature $t=\ln\Lambda/T$, we expect a linear behavior for the derivatives of $\ln^2$-divergent bubbles, i.e. $\chi^0_{\mathrm{pp}}$ and $\chi^M_{\mathrm{ph}}$,  and a constant for the derivatives of $\ln$-divergent bubbles, i.e. $\chi^M_{\mathrm{pp}}$ and $\chi^0_{\mathrm{ph}}$ in the nested case.
Furthermore, a finite $t_3$ destroys nesting at the lowest scales, so that the growth of $\chi^M_{\mathrm{ph}}$ is stopped.
In the case with the eVHS, we expect the particle-particle bubble with zero incoming momentum and the particle-hole bubble with zero momentum transfer to grow like a power law, which is an exponential when plotted against $t=\ln\Lambda/T$.
This is exactly what we find, in agreement to the approximate analytical calculation Eqs.~\eqref{eq:chipp0} and \eqref{eq:chiph0}.
This confirms that there is a qualitative change in the loop hierarchy with  $\chi^0_{\mathrm{pp}}$ and  $\chi^0_{\mathrm{ph}}$ growing large at small temperatures while the other loops become subleading.
In Fig.~\ref{fig:dtBt3}, we show how this hierarchy changes when $t_3$ is varied.
We can also extract an estimate for the parameter $d_0$ when we compare the magnitude of $\dot \chi^0_{\mathrm{pp}}$ and $\dot \chi^0_{\mathrm{ph}}$.
We obtain a similar value as before $d_0\approx0.30$.

\section{RG vs. perturbation expansion}
\label{sec:2loopdiagram}

To verify the accuracy of the RG equations, it is instructive to compare it with a direct perturbative computation of the renormalization of the couplings.  A straightforward comparison shows that our RG approach reproduces the correct temperature dependence and the prefactors of the contributions coming from the ladder diagrams, and that it also contains contributions, which in perturbation theory come from the diagrams which contain segments with particle-particle and particle-hole bubbles.
However, these last contributions are not reproduced exactly within our RG.  Specifically, the power-law forms of the temperature dependencies  are captured correctly, but the prefactors are reproduced only up to corrections of order one.
In the limit of $\kappa\rightarrow 0$, i.e. for the logarithmic RG, these deviations vanish, i.e., to logarithmic accuracy, RG exactly reproduces perturbation theory order by order.

\begin{figure}[t]
\begin{center}
\includegraphics[width=0.9\columnwidth]{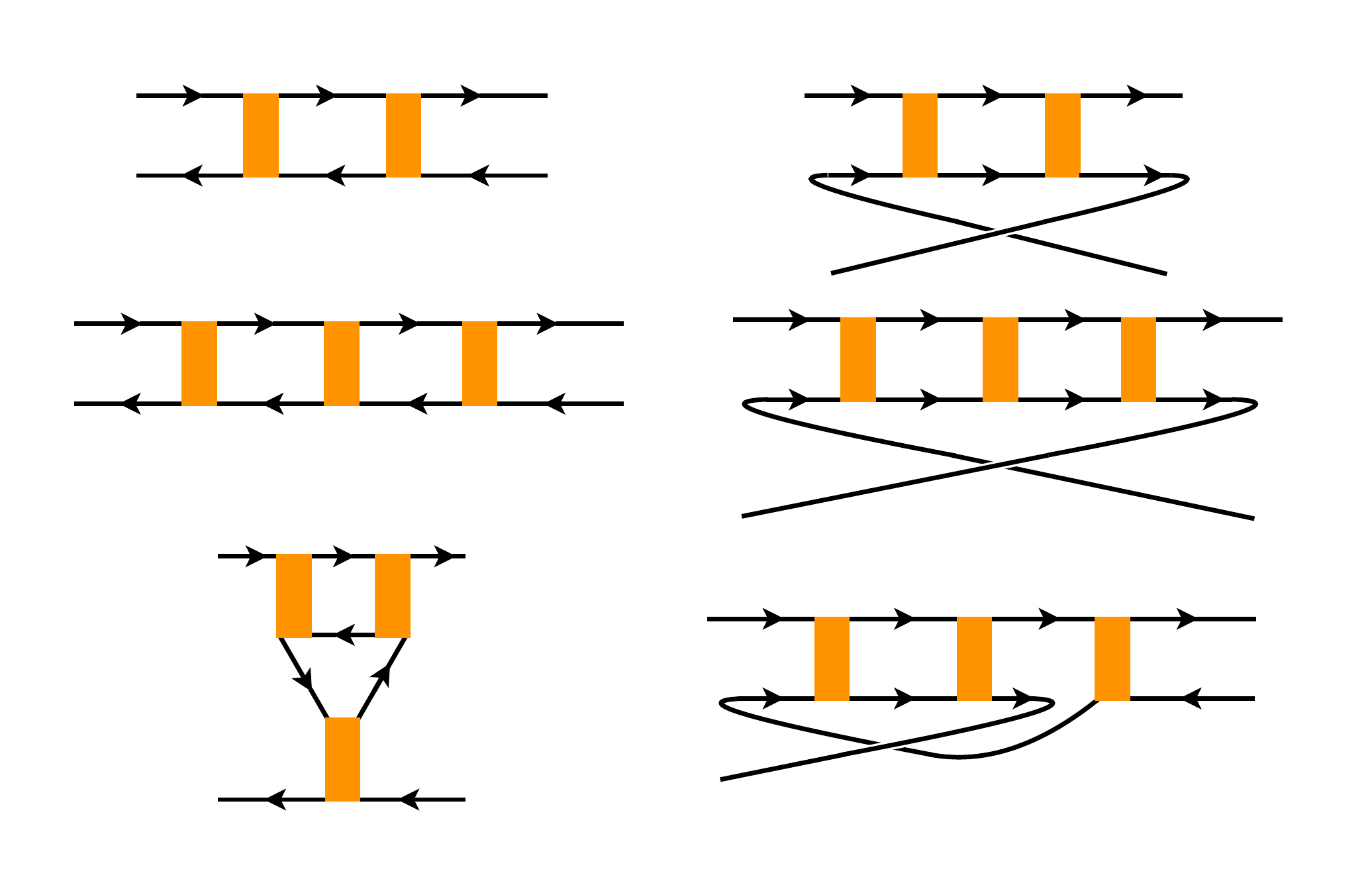}
\end{center}
\caption{\textbf{Exemplary diagrams} at one and two-loop level if $g_1=g_2=g_3=0$. Diagrams can be classified into three channels, often denoted as crossed particle-hole, particle-particle, and direct particle-hole channel.
On the one-loop level (first line), we have contributions from the crossed particle-hole and the particle-particle channel. Diagrams in the direct particle-hole channel cancel against each other.
On the two-loop level, we can distinguish pure diagrams, which consist only of contributions belonging to the same channel, and mixed diagrams, which contain sub-diagrams from different channels. The second line contains pure diagrams in the crossed particle-hole and the particle-particle channel. The last line shows examples of mixed diagrams: a mixed crossed and direct particle-hole (left) and a mixed particle-particle and particle-hole diagram (right).  The RG approximates the mixed diagrams as the product of its subdiagrams.}
\label{fig:2loop}
\end{figure}

To exemplify this, we compare the expressions, obtained from the RG and from the direct  perturbation expansion up to two-loop order.
For clarity, we simplify the problem and set $g_1=g_2=g_3=0$, i.e., consider the case when only the $g_4$ coupling is non-zero (with our RG, if bare $g_1=g_2=g_3=0$, the dressed couplings also vanish).
Then, the RG equation for $g_4$ is
\begin{equation}
\partial_b g_4=(d_0-1)g_4^2\,,
\end{equation}
cf. Eq.~\eqref{betag4b}.
Solving this equation iteratively, starting from bare $g_0$, we obtain the following series
\begin{equation}
g_4=g_{0} + (d_0-1) b g_{0}^2 +(d_0^2-2d_0+1) b^2 g_{0}^3.
\label{eq:rg2loop}
\end{equation}
Within the diagrammatic perturbation theory, the $ O(b g_{0}^2)$ term comes from the one-loop diagrams, and the
$ O(b^2 g_{0}^3)$ term comes from the two-loop diagrams.  At one-loop order, the two diagrams describe the renormalization of $g^4$ by particle-hole and particle-particle bubbles (see Fig.~\ref{fig:2loop}). There are $b g^2_0$ contributions from other diagrams (not shown), but they cancel out. Evaluating these diagrams, we reproduce the prefactor $d_0-1$ in (\ref{eq:rg2loop}).

We next move to two-loop order. One can identify what kinds of two-loop diagrams would reproduce the three terms of order $b^2 g^3_0$ in Eq. (\ref{eq:rg2loop}). 
The power of $d_0$ indicates
that the term $ d_0^2b^2 g^3_0$ is the contribution with two particle-hole bubbles
\begin{align}
\label{phRG}
D_{\mathrm{ph}}^{\mathrm{RG}}
&=T^2 g^3_0 \sum_{k_0,q_0}\int d\vec k \int d\vec q G^2(q) G^2(k)\,.
\end{align}
For brevity, we collect Matsubara frequency and momentum as $k=(k_0,\vec k)$. 
Here and below we assume that $\vec q$ and $\vec k$ denote deviations from the $M_i$ point.
The term $-2d_0b^2 g^3_0$ is a mixed particle-particle/particle-hole contribution
\begin{align}
D_{\mathrm{pp,ph}}^{\mathrm{RG}}
&=2T^2 g^3_0 \sum_{k_0,q_0}\int d\vec k \int d\vec q G^2(q) G(k)G(-k)\,,
\end{align}
and the term $b^2 g^3_0$ is the contribution  from two bubbles in the particle-particle channel:
\begin{align}
\label{ppRG}
D_{\mathrm{pp}}^{\mathrm{RG}}
&=T^2 g^3_0 \sum_{k_0,q_0}\int d\vec k \int d\vec q G(q)G(-q) G(k)G(-k).
\end{align}
Here and below we use that $-\vec k -M_i=-\vec k+M_i$ (up to reciprocal lattice vector), i.e. $\vec q$ and $\vec k$ also denote deviations from the $M_i$ point in particle-particle bubbles.
We see that in all contributions in Eqs.~\eqref{phRG}-\eqref{ppRG} the integration/summation over $k$ and $q$ decouples, i.e., these terms are the products of one-loop diagrams.
%

 In the  direct perturbation theory,
 the perturbative contributions can also be assembled into contributions from two particle-hole 
 loops, two particle-particle 
 loops, and one particle-hole and one particle-particle loop. 
   Exemplary diagrams (not all),  which contribute to the renormalization of $g_4$ at two-loop order, are shown in Fig.~\ref{fig:2loop}.
   
The two-loop diagram with two particle-particle loops (top right one in Fig.~\ref{fig:2loop}) is the direct product of two one-loop particle-particle diagrams, and it yields the same result as in the RG:
\begin{align}
 D_{\mathrm{pp}}^{\mathrm{pert}}&=g_0^3T^2 \sum_{k_0,q_0}\int\!d\vec k\! \int\!d\vec q\ G(q)G(-q) G(-k)G(k)\nonumber \\
&=D_{\mathrm{pp}}^{\mathrm{RG}}\,,
\end{align}
In the particle-hole channel, we get two types of diagrams from the perturbation expansion. Diagrams of one type are the ones in which the integration/summation over $k$ and $q$ factorizes, ( e.g. the top left diagram in Fig.~\ref{fig:2loop}).
In the diagrams of the other type, (e.g. the bottom left one in Fig.~\ref{fig:2loop}), there is no factorization.
In total, the  two-loop particle-hole diagrams sum to
\begin{align}
D_{\mathrm{ph}}^{\mathrm{pert}}&=2T^2 g^3_0 \sum_{k_0,q_0}\int\!d\vec k\! \int\!d\vec q\ G^2(q) G^2(k) \nonumber \\
&-2T^2 g^3_0 \!\sum_{k_0,q_0}\int\!d\vec k \int\!d\vec q\ G^2(q) G(k)G(M_i+k-q)\,.\nonumber
\label{a}
\end{align}
Here and below we have set all external frequencies and momenta to be the same, assuming they are located at $M$ points $M_i$.
In the logarithmic case with DOS $\rho(\epsilon)=\rho_0\ln(\Lambda/\epsilon)$, the main contribution in the coupled integral comes from $\epsilon_k\gg \epsilon_q$ and we can approximate the second integral via
\begin{align}
&T^2\!\sum_{k_0,q_0}\int\!d\vec k \int\!d\vec q\ G^2(q) G(k)G(M_i+k-q)\notag\\
\approx& T^2\!\sum_{k_0,q_0} \int\!d\vec q\ G^2(q) \int_{\epsilon_k\geq \epsilon_q}\!d\vec k \,G^2(k)\notag\\
=& \int d\epsilon \rho(\epsilon) n'_F(\epsilon)  \int_{\epsilon'>\epsilon} d\epsilon' \rho(\epsilon') n'_F(\epsilon')\notag\\
\approx&\rho_0^2 \ln^2(\Lambda/T) \int d\epsilon \frac{1}{4\cosh^2(\epsilon/2)}  \int_{\epsilon'>\epsilon} d\epsilon'  \frac{1}{4\cosh^2(\epsilon'/2)} \notag\\
=&\frac{1}{2} \left( \rho_0 \ln(\Lambda/T) \int d\epsilon \frac{1}{4\cosh^2(\epsilon/2)}  \right)^2\notag \\
=& \frac{1}{2}T^2\!\sum_{k_0,q_0} \int\!d\vec q \int\!d\vec k \ G^2(q) G^2(k)
\end{align}
Here, we have neglected terms that are smaller than $\ln^2(\Lambda/T)$. Substituting into \ref{a}),  we find that the direct perturbation theory yields, to logarithmic accuracy,
\beq
D_{\mathrm{ph}}^{\mathrm{pert}} = D_{\mathrm{ph}}^{\mathrm{RG}} 
\eeq
However, in the case of HOVh point, we cannot decouple the integrations over ${\bf k}$ and ${\bf q}$, i.e. $D_{\mathrm{ph}}^{\mathrm{pert}}\neq D_{\mathrm{ph}}^{\mathrm{RG}}$. Thus, there is a difference between the RG and the perturbation expansion. 

We show below that this leads to corrections in the prefactor, while the temperature dependence is reproduced correctly, i.e. $D_{\mathrm{ph}}^{\mathrm{pert}}= c D_{\mathrm{ph}}^{\mathrm{RG}}$ with a constant $c=\mathcal O(1)$.

Finally, the mixed particle-particle/particle-hole diagrams in the perturbation expansion 
sum to
\begin{align}
&D_{\mathrm{pp,ph}}^{\mathrm{pert}}=2T^2g_0^3\sum_{k_0,q_0}\int d\vec k \int d\vec q G^2(q)G(k)G(M_i+q-k)\nonumber\\
&+2T^2 g_0^3\!\sum_{k_0,q_0}\int\!d\vec k\!\int\!d\vec q G(q)G(-q)G(k)G(M_i+k+q)\,.
\label{ppph2}
\end{align}
(see bottom right diagram in Fig.~\ref{fig:2loop} for an example).
The second term in (\ref{ppph2}) becomes subleading in the logarithmic case, because the second logarithm of the particle-particle channel is cut by the particle-hole insertion:
\begin{align}
&T^2\!\sum_{k_0,q_0}\int\!d\vec q \int\!d\vec k\!G(q)G(-q)G(k)G(M_i+k+q)\notag\\
\approx&T^2\!\sum_{k_0,q_0}\!\int\!d\vec q \int_{\epsilon_k\geq\epsilon_q}\!d\vec k\, G(q)G(-q)G^2(k)\notag\\
=& \int d\epsilon \rho(\epsilon) \frac{1-2n_F(\epsilon)}{2\epsilon}  \int_{\epsilon'>\epsilon} d\epsilon' \rho(\epsilon') n'_F(\epsilon')\notag\\
\approx&-\rho_0^2\ln^2(\Lambda/T) \int d\epsilon\frac{\tanh{\epsilon/2}}{2\epsilon}\int_{\epsilon'>\epsilon} d\epsilon'\frac{1}{4\cosh^2\epsilon'/2}\notag\\
=&\rho_0^2\ln^2(\Lambda/T) \int d\epsilon\frac{\tanh{\epsilon/2}}{2\epsilon}\,\frac{\tanh\epsilon/2-1}{2}
\end{align}
This contribution is of order $\ln^2(\Lambda/T)$. In contrast, the first term in (\ref{ppph2}) is of order $\ln^3(\Lambda/T)$. Indeed.
\begin{align}
&T^2\sum_{k_0,q_0}\int d\vec k \int d\vec q G^2(q)G(k)G(M_i+q-k)\nonumber\\
\approx& T^2\sum_{k_0,q_0} \int d\vec q \int_{\epsilon_k\geq\epsilon_q} d\vec kG^2(q)G(k)G(-k)\nonumber\\
=& \int d\epsilon \rho(\epsilon)  n'_F(\epsilon)  \int_{\epsilon}^{\Lambda/T} \!d\epsilon' \rho(\epsilon')\frac{1-2n_F(\epsilon')}{2\epsilon'}\notag\\
\approx&-\rho_0^2\ln^2(\Lambda/T) \int d\epsilon \frac{1}{4\cosh^2\epsilon/2} \int_{\epsilon}^{\Lambda/T} d\epsilon' \frac{\tanh{\epsilon'/2}}{2\epsilon'} \notag\\
 =& O(\ln^3(\Lambda/T)).
\end{align}
One can verify that to logarithmic accuracy,
\begin{align}
&T^2\sum_{k_0,q_0}\int d\vec k \int d\vec q G^2(q)G(k)G(M_i+q-k) \notag\\
=& T^2\sum_{k_0,q_0} \int d\vec q \int d\vec kG^2(q)G(k)G(-k).
\end{align}
We note that if the calculation is performed at zero temperature and regularized by a non-zero deviation from the Van Hove points, both contributions in Eq.~\ref{ppph2} are of order $\ln^3(\Lambda/T)$, but eventually sum to the same prefactor as the one at a non-zero temperature.
So we find that for the logarithmic DOS (the case  $\kappa\rightarrow 0$)
\beq
D_{\mathrm{pp,ph}}^{\mathrm{pert}}= D_{\mathrm{pp,ph}}^{\mathrm{RG}}
\eeq
to logarithmic accuracy.
For our case of HOVH points, $D_{\mathrm{pp,ph}}^{\mathrm{pert}}$ and $D_{\mathrm{pp,ph}}^{\mathrm{RG}}$ are not equivalent because the integrals over ${\bf k}$ and ${\bf q}$ do not decouple. Like before, $D_{\mathrm{pp,ph}}^{\mathrm{pert}}= \tilde c D_{\mathrm{pp,ph}} $ with $\tilde c=\mathcal O(1)$.
For verification, we explicitly calculate the coupled integrals to show that we still get the correct temperature dependence.
\begin{widetext}
For concreteness, we consider Van Hove filling and $\kappa=1/4$.
We use the dispersions for the vicinity of the $M$ points Eq.~\eqref{eq:m1approx_1}, i.e. $\epsilon_{\vec q}=\epsilon_{M_1}(\vec q)$, $\epsilon_{\vec k}=\epsilon_{M_1}(\vec k)$ and $\epsilon_{M_i+\vec q-\vec k}=\epsilon_{M_1}(\vec q-\vec k)$ (we also abbreviate $\epsilon_{M_1}(\vec p)$ by $\epsilon^{M_1}_{\vec p}$). We denote the external frequency by $ip_0$.
For Matsubara sums, we use
\begin{align}
T\sum_{i\omega}\frac{1}{(i\omega-\epsilon_1)(i\omega-\epsilon_2)}&=\frac{n_F(\epsilon_1)-n_F(\epsilon_2)}{\epsilon_1-\epsilon_2}\\[8pt]
T\sum_{i\omega}\frac{1}{(i\omega-\epsilon_1)(i\omega-\epsilon_2)(i\omega-\epsilon_3)}&=\frac{n_F(\epsilon_1)}{(\epsilon_1-\epsilon_2)(\epsilon_1-\epsilon_3)}-\frac{n_F(\epsilon_2)}{(\epsilon_1-\epsilon_2)(\epsilon_2-\epsilon_3)}+\frac{n_F(\epsilon_3)}{(\epsilon_1-\epsilon_3)(\epsilon_2-\epsilon_3)}\\[8pt]
T\sum_{i\omega}\frac{1}{(i\omega-\epsilon_1)^2(i\omega-\epsilon_2)}&=\frac{n_F'(\epsilon_1)}{\epsilon_1-\epsilon_2}-\frac{n_F(\epsilon_1)}{(\epsilon_1-\epsilon_2)^2}+\frac{n_F(\epsilon_2)}{(\epsilon_1-\epsilon_2)^2}
\end{align}
and $n_F(ip_0+\epsilon)=-n_B(\epsilon)$, $n_B(\epsilon_2-\epsilon_1)\left[ n_F(\epsilon_1)-n_F(\epsilon_2) \right]=n_F(-\epsilon_1)n_F(\epsilon_2)$, where $n_F$ is the Fermi and $n_B$ the Bose function.
We obtain for the coupled particle-particle/particle-hole diagram
\begin{align}
&2T^2\sum_{k_0,q_0}\int d\vec k \int d\vec q \frac{1}{(iq_0-\epsilon^{M_1}_{\vec q})^2(ik_0-\epsilon^{M_1}_{\vec k})(-ik_0+iq_0+ip_0-\epsilon^{M_1}_{\vec q-\vec k})}\nonumber\\
=&2\int d\vec k\int d\vec q\Bigg(\left[1- n_F(\epsilon^{M_1}_{\vec q-\vec k}) - n_F(\epsilon^{M_1}_{\vec k})  \right] \left[ \frac{- n'_F(\epsilon^{M_1}_{\vec q}) }{(ip_0+\epsilon^{M_1}_{\vec q}-\epsilon^{M_1}_{\vec k}+\epsilon^{M_1}_{\vec q-\vec k})} +\frac{ n_F(\epsilon^{M_1}_{\vec q}) }{(ip_0+\epsilon^{M_1}_{\vec q}-\epsilon^{M_1}_{\vec k}+\epsilon^{M_1}_{\vec q-\vec k})^2} \right]\nonumber \\
&\hspace{2cm}+\frac{ n_F(\epsilon^{M_1}_{\vec k}) n_F(\epsilon^{M_1}_{\vec k-\vec q})}{(ip_0+\epsilon^{M_1}_{\vec q}-\epsilon^{M_1}_{\vec k}+\epsilon^{M_1}_{\vec q-\vec k})^2}\Bigg)
\end{align}
To see the temperature dependence of this expression, we rescale $k_x=T^{1/4}\tilde k_x, k_y=\sqrt{T}\tilde k_y,q_x=T^{1/4}\tilde q_x, q_y=\sqrt{T}\tilde q_y$ and express the external frequency as $p_0=(2n+1)i\pi T$, which yields
\begin{align}
\frac{2}{\sqrt{T}}\int d\vec{\tilde k}\int d\vec{\tilde q}&\Bigg(\left[1- \tilde n_F(\epsilon^{M_1}_{ \tilde q-\tilde  k}) - n_F(\epsilon^{M_1}_{\tilde  k})  \right] \left[ \frac{[4\cosh^2(\epsilon^{M_1}_{\vec q})]^{-1} }{((2n+1)i\pi+\epsilon^{M_1}_{\tilde  q}-\epsilon^{M_1}_{\tilde  k}+\epsilon^{M_1}_{\tilde  q-\tilde k})} +\frac{ \tilde n_F(\epsilon^{M_1}_{\tilde  q}) }{((2n+1)i\pi+\epsilon^{M_1}_{\tilde q}-\epsilon^{M_1}_{\tilde k}+\epsilon^{M_1}_{\tilde q-\tilde k})^2} \right]\nonumber \\
&+\frac{ \tilde n_F(\epsilon^{M_1}_{\tilde  k}) \tilde n_F(\epsilon^{M_1}_{\tilde k-\tilde q})}{((2n+1)i\pi+\epsilon^{M_1}_{\tilde q}-\epsilon^{M_1}_{\tilde k}+\epsilon^{M_1}_{\tilde q-\tilde k})^2}\Bigg)
\end{align}
where we defined $\tilde n_F(x)=1/(1+\exp(x))$.
The integrand is finite and independent of temperature, so indeed, we obtain the correct temperature dependence $T^{2\kappa}$ for $\kappa=1/4$. The same is true for the second integral in Eq.~\eqref{ppph2}.
In the particle-hole channel, we get
\begin{align}
&-2T^2\sum_{k_0,q_0}\int d\vec k \int d\vec q \frac{1}{(iq_0-\epsilon^{M_1}_{\vec q})^2(ik_0-\epsilon^{M_1}_{\vec k})(ik_0-iq_0+ip_0-\epsilon^{M_1}_{\vec k-\vec q})}\nonumber \\[5pt]
=&-2\int d\vec k \int d\vec q \left[ \left( n_F(\epsilon^{M_1}_{\vec k}) -n_F(\epsilon^{M_1}_{\vec k-\vec q}) \right) \left( \frac{n'_F(\epsilon^{M_1}_{\vec q})}{ip_0+\epsilon^{M_1}_{\vec q}+\epsilon^{M_1}_{\vec k}-\epsilon^{M_1}_{\vec k-\vec q}} -\frac{n_F(\epsilon^{M_1}_{\vec q})}{(ip_0+\epsilon^{M_1}_{\vec q}+\epsilon^{M_1}_{\vec k}-\epsilon^{M_1}_{\vec k-\vec q})^2} \right)\right. \nonumber \\
&\hspace{2.7cm}\left. - n_F(\epsilon^{M_1}_{\vec k -\vec q}) \frac{1-2n_F(\epsilon^{M_1}_{\vec k})}{(ip_0+\epsilon^{M_1}_{\vec q}+\epsilon^{M_1}_{\vec k}-\epsilon^{M_1}_{\vec k-\vec q})^2}\right]\,,
\end{align}
%
\end{widetext}
which after rescaling is also proportional to $1/\sqrt{T}$. We can calculate the remaining integrals numerically.
The calculation does
indeed
show that there is a factor of order $\mathcal O(1)$ difference between the perturbation expansion and the RG. These results can be generalized to arbitrary loop order with the result that the iterative solution of the RG equations reproduces the temperature dependence, which one obtained in the order-by-order diagrammatic expansion, and the prefactors are generally different because for power-law-singular DOS the momentum integrations do not factorize.

We note that the discrepancy between the perturbation expansion and the RG may be systematically studied by employing advanced truncation schemes such as, e.g., the recently developed multi-loop functional RG~\cite{PhysRevLett.120.057403}.

\section{Three-patch RG from functional RG}
\label{sec:rg}

To investigate the quantum many-body instabilities of our model we employ a parquet renormalization group (RG) approach.
The parquet RG flow equations can be straightforwardly derived within a more general functional integral framework based on a one-loop exact functional renormalization group (fRG) flow equation for the one-particle irreducible vertices of a correlated fermion system, see Refs.~\onlinecite{RevModPhys.84.299,platt2013functional,Dupuis:2020fhh} for reviews.

With this renormalization group scheme we can then identify the leading instabilities in the presence of competing interactions by successively integrating out fermion degrees of freedom starting from an initial RG scale $\Lambda_0$, e.g., corresponding to the bandwidth down to the infrared scale $\Lambda \to 0$. We now briefly set up the functional RG approach.

We consider the action for a  many-fermion system corresponding to our model Hamiltonian, i.e.
\begin{align}
	S[\bar\psi,\psi]=-(\bar\psi,G_0^{-1}\psi)+V[\bar\psi,\psi]\,,
\end{align}
where $\bar\psi,\psi$ are the Grassmann-valued fermion field degrees of freedom, the first term is the quadratic part including the free fermion propagator $G_0(\omega_n,\vec{k},b)=1/(i\omega_n-\epsilon_b(\vec{k}))$, the Matsubara frequency $\omega_n$ and wavevector $\vec{k}$. The energy dispersion $\epsilon_b(\vec{k})$ with band index $b$ follows from diagonalization of the free part of the Hamiltonian $H_0$ and we assume that the fermionic propagator is diagonal with respect to the spin quantum number.
The second term $V[\bar\psi,\psi]$ in the above equation is the interaction term, which is quartic in the fermionic fields $\bar\psi,\psi$ and can be inferred from the interaction part of the Hamiltonian.

To set up the functional RG flow equations, the bare propagator is regularized  by an infrared momentum cutoff represented by the scale $\Lambda$,
\begin{align}
	G_0(\omega_n,\vec{k},b)\to G_0^\Lambda(\omega_n,\vec{k},b)\,.
\end{align}
The purpose of the regularization and the introduction of the modified propagator $G_0^\Lambda$ is to cut of infrared modes below the scale $\Lambda$ and the implementation of this regularization can be realized in different ways, i.e. employing a momentum cutoff, a frequency cutoff, or a temperature cutoff. We leave this choice open for the moment, as it does not affect the structure of the fRG equations.

The modified propagator $G_0^\Lambda$ is now used in the generating functional for the one-particle irreducible correlation functions and an exact flow equation is generated upon variation with respect to the cutoff scale $\Lambda$. More explicitly, we start with the generating functional for the fully connected correlation functions~\cite{negele2018quantum}
\begin{align}
	\mathcal{G}[\bar\eta,\eta]=-\ln \int\mathcal{D}\psi\mathcal{D}\bar\psi\ e^{-S[\bar\psi,\psi]+(\bar\eta,\psi)+(\bar\psi,\eta)}\,.
\end{align}
For convenience, we consider the Legendre transform of $\mathcal{G}[\bar\eta,\eta]$, i.e. $\Gamma[\bar\psi,\psi]=(\bar\eta,\psi)+(\bar\psi,\eta)+\mathcal{G}[\bar\eta,\eta]$, which is called the effective action and which generates the one-particle irreducible correlation functions. Note that the field arguments in the effective action $\Gamma$ are $\psi=-\partial\mathcal{G}/\partial\bar\eta$ and  $\bar\psi=\partial\mathcal{G}/\partial\eta$.

Using the modified propagator $G_0^\Lambda$ provides a cutoff dependence to the effective action $\Gamma \to \Gamma^\Lambda$. Taking the derivative of that scale-dependent effective action with respect to $\Lambda$ produces an exact RG flow equation, reading
\begin{align}
	\frac{\partial}{\partial \Lambda}\Gamma^\Lambda[\bar\psi,\psi]=&-(\bar\psi,(\dot{G}^\Lambda_0)^{-1}\psi)\notag\\
	&-\frac{1}{2}\mathrm{Tr}\left((\dot{\mathbf{G}}^\Lambda_0)^{-1}
	(\dot{\mathbf{\Gamma}}^{(2)\Lambda}[\bar\psi,\psi])^{-1}
	\right)\,,\label{eq:exRG}
\end{align}
where $(\dot{\mathbf{G}}^\Lambda_0)^{-1}=\mathrm{diag}((G^\Lambda_0)^{-1},(G^{\Lambda t}_0)^{-1})$ and
\begin{align}
	\mathbf{\Gamma}^{(2)\Lambda}[\bar\psi,\psi]=
	\begin{pmatrix}
		\frac{\partial^2\Gamma^\Lambda}{\partial\bar\psi\partial\psi} & \frac{\partial^2\Gamma^\Lambda}{\partial\bar\psi\partial\bar\psi}\\
		\frac{\partial^2\Gamma^\Lambda}{\partial\psi\partial\psi} & \frac{\partial^2\Gamma^\Lambda}{\partial\psi\partial\bar\psi}
	\end{pmatrix}
\end{align}
The inital condition of this differential equation is defined at the ultraviolet scale $\Lambda_\mathrm{UV}$ by the microscopic action $\Gamma^{\Lambda_\mathrm{UV}}=S$ and in the limit $\Lambda\to 0$ by the full quantum effective action $\Gamma$.

For tractability of the exact flow equation, we employ a truncation of the effective action $\Gamma^\Lambda$ in form of the vertex expansion ansatz
\begin{align}
	\Gamma^\Lambda[\psi,\bar\psi]=&\sum_{i=0}^\infty \frac{(-1)^i}{(i!)^2}\sum_{\substack{k_1,...,k_i\\ k_1^\prime,..., k_i^\prime}}\Gamma^{(2i)\Lambda}(k_1^\prime,..., k_i^\prime,k_1,...,k_i)\nonumber\\
	&\times \bar\psi(k_1^\prime)...\bar\psi(k_i^\prime)\psi(k_i)...\psi(k_1)\,.
\end{align}
This ansatz is inserted into the exact flow equation, which generates a hierarchy of flow equations for the one-particle irreducible vertex functions $\Gamma^{(2i)\Lambda}$.
We truncate the tower of flow equations and exclusively consider the RG evolution of the two-particle interaction $\Gamma^{(4)\Lambda}$, which carries spin indices $\sigma_i$ and multi-indices $k$ collecting Matsubara frequencies, wave-vectors and band indices. We also neglect the self-energy feedback.

For the spin-rotation invariant system, we consider in this work, the two-particle interaction can be written as
\begin{align}
	\Gamma^{(4)\Lambda}_{\sigma_1\sigma_2\sigma_3\sigma_4}=V^\Lambda\delta_{\sigma_1\sigma_3}\delta_{\sigma_2\sigma_4}-V^\Lambda\delta_{\sigma_1\sigma_4}\delta_{\sigma_2\sigma_3}\,,
\end{align}
introducing the effectice interaction vertex $V^\Lambda=V^\Lambda(k_1,k_2,k_3,b_4)$. For the analysis of instabilities, we are interested in the most singular part of $V^\Lambda$, which comes from the smallest Matsubara frequency and we therefore only consider this one.
Then, the RG flow of $V^\Lambda$ can be derived from the exact flow equation Eq.~\eqref{eq:exRG} and reads
\begin{align}\label{eq:vertexfrg}
	\frac{d}{d\Lambda}V^\Lambda=\tau_\mathrm{pp}+\tau_\mathrm{ph,d}+\tau_\mathrm{ph,cr}\,.
\end{align}
with the particle-particle, the direct paricle-hole, and the crossed particle-hole contributions on the right hand side of the equation, reading
\begin{align}
	\tau_{\mathrm{pp}}=-\frac{1}{2}\SumInt V^\Lambda(k_1,k_2,k,b^\prime)L^\Lambda(k,q_{\mathrm{pp}})V^\Lambda(k,q_{\mathrm{pp}},k_3,b_4)\,,\notag
\end{align}
where we defined $\SumInt=-A^{-1}_{\mathrm{BZ}}T\sum_\omega\int d^2 k \sum_{b,b^\prime}$ and $A_{\mathrm{BZ}}$ is the area of the Brillouin zone.
Further, we have
\begin{align}
	\tau_{\mathrm{ph,d}}=&\frac{1}{2}\SumInt [2V^\Lambda(k_1,k,k_3,b^\prime)L^\Lambda(k,q_{\mathrm{d}})V^\Lambda(q_{\mathrm{d}},k_2,k,b_4)\notag\\	
&-V^\Lambda(k,k_1,k_3,b^\prime)L^\Lambda(k,q_{\mathrm{d}})V^\Lambda(q_{\mathrm{d}},k_2,k,b_4)\notag\\
&-V^\Lambda(k,k_1,k_3,b^\prime)L^\Lambda(k,q_{\mathrm{d}})V^\Lambda(k_2,q_{\mathrm{d}},k,b_4)]\,,\notag
\end{align}
and
\begin{align}
	\tau_{\mathrm{ph,cr}}=-\frac{1}{2}\SumInt V^\Lambda(k,k_2,k_3,b^\prime)L^\Lambda(k,q_{\mathrm{cr}})V^\Lambda(k_1,q_{\mathrm{cr}},k,b_4)\,.\notag
\end{align}
Above, we have used the definitions $q_{\mathrm{pp}}=-k+k_1+k_2$, $q_{\mathrm{d}}=k+k_1-k_3$ and $q_{\mathrm{cr}}=k+k_2-k_3$ and the loop kernel
\begin{align}
L^\Lambda(k,k^\prime)=\frac{d}{d\Lambda}[G_0^\Lambda(k)G^\Lambda_0(k^\prime)]\,,
\end{align}
with the free modified propagator $G_0^\Lambda$ due to the neglect of the self-energy.

\begin{figure*}[t!]
\includegraphics[width=0.9\textwidth]{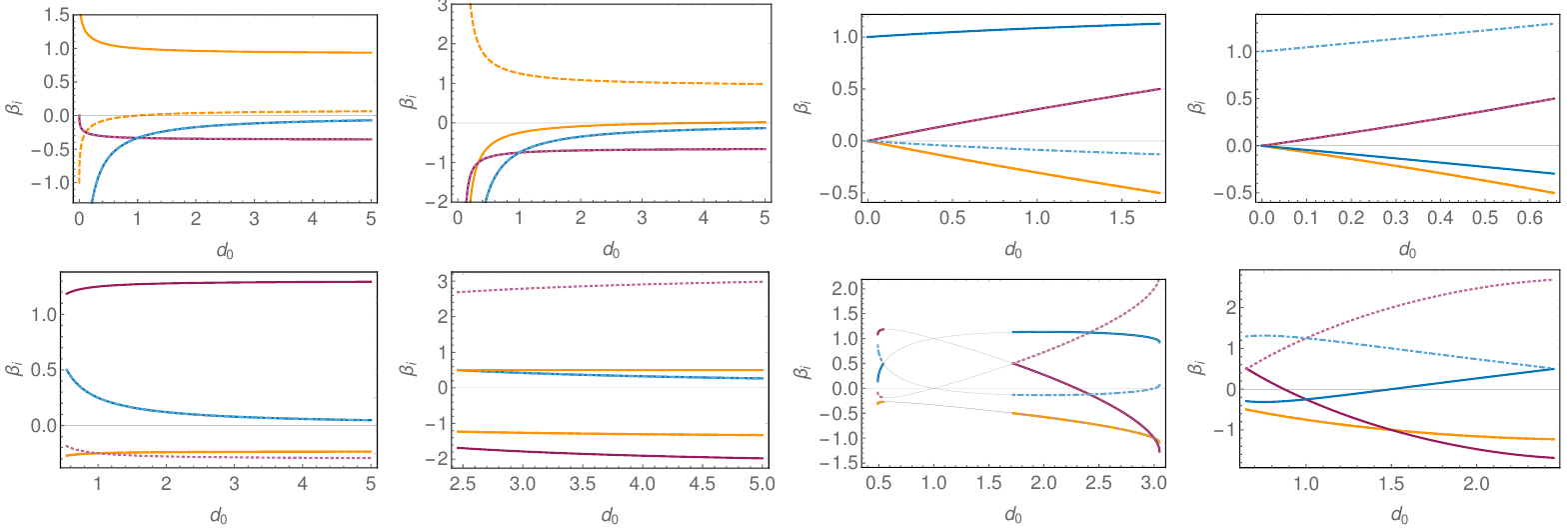}
\caption{\textbf{Susceptibility exponents} for the stable fixed trajectories as function of $d_0$. They correspond to (I)-(VIII) from left to right, from top to bottom. Color code: s-wave spin (solid, orange), d-wave spin (dashed, light orange), s-wave charge (solid, purple), d-wave charge (dotted, light purple), s-wave SC (solid, blue), d-wave SC (dot-dashed, light blue). Fixed trajectory (VII) has two regimes of stability, for clarity we connect the exponents in both regimes by thin lines.}
\label{fig:exponents}
\end{figure*}

To derive the $N=3$-patch parquet RG flow equations, cf. Eqs.~\eqref{betag1start} -- \eqref{betag4start}, we now introduce further approximations. Since we are interested in instabilities, we will consider only the strongest contributions to the flow of $V^\Lambda$, which come from wave-vectors where the density of states is large, i.e. in our model the wave-vectors at/near the $M_{1,2,3}$ points.
Therefore, we introduce an $N=3$ patch approximation by evaluating $V^\Lambda$ only at the singularity momenta $M_{1,2,3}$.
We exclusively take into account two-particle scattering processes on these three patches, as indicated in Fig.~\ref{fig:gology}.
Further, we assume that the interaction vertex is approximately constant within small patches surrounding the $M_i$ points where the energy dispersion can be approximated by a pure saddle point behavior.
We can thus relate the interaction vertex $V^\Lambda$ with the interaction couplings $g_i$, cf. Fig.~\ref{fig:gology},
\begin{align}
	V^\Lambda(M_i,M_j,M_i,M_j)&=g_1,\quad i\neq j\\
	V^\Lambda(M_i,M_j,M_j,M_i)&=g_2,\quad i\neq j\\
	V^\Lambda(M_i,M_i,M_j,M_j)&=g_3,\quad i\neq j\\
	V^\Lambda(M_i,M_i,M_i,M_i)&=g_4\,.
\end{align}
These relations can be put into the flow equation for the interaction vertex, Eq.~\ref{eq:vertexfrg}, yielding,
%
\begin{widetext}
%
\begin{align}
	\frac{d}{d\Lambda}g_1&=\frac{d}{d\Lambda}V^\Lambda(M_1,M_2,M_1,M_2)=\tau_\mathrm{pp}(M_1,M_2,M_1,M_2)+\tau_\mathrm{ph,d}(M_1,M_2,M_1,M_2)+\tau_\mathrm{ph,cr}(M_1,M_2,M_1,M_2)\notag\,,\\
	\frac{d}{d\Lambda}g_2&=\frac{d}{d\Lambda}V^\Lambda(M_1,M_2,M_2,M_1)=\tau_\mathrm{pp}(M_1,M_2,M_2,M_1)+\tau_\mathrm{ph,d}(M_1,M_2,M_2,M_1)+\tau_\mathrm{ph,cr}(M_1,M_2,M_2,M_1)\notag\,,\\
	\frac{d}{d\Lambda}g_3&=\frac{d}{d\Lambda}V^\Lambda(M_1,M_1,M_2,M_2)=\tau_\mathrm{pp}(M_1,M_1,M_2,M_2)+\tau_\mathrm{ph,d}(M_1,M_1,M_2,M_2)+\tau_\mathrm{ph,cr}(M_1,M_1,M_2,M_2)\notag\,,\\
	\frac{d}{d\Lambda}g_4&=\frac{d}{d\Lambda}V^\Lambda(M_1,M_1,M_1,M_1)=\tau_\mathrm{pp}(M_1,M_1,M_1,M_1)+\tau_\mathrm{ph,d}(M_1,M_1,M_1,M_1)+\tau_\mathrm{ph,cr}(M_1,M_1,M_1,M_1)\notag\,.
\end{align}
%
\end{widetext}
%
Evaluating the various channels contributions within the small patches around the $M$ points and for the respective wave-vector configurations  then -- after some straightforward algebra -- yields the flow equations for the interaction couplings $g_i, i\in \{1,2,3,4\}$ presented in the main text, cf. Eqs.~\eqref{betag1start} -- \eqref{betag4start}.
%

\section{RG fixed points}
\label{sec:fps}
We look for fixed points in the flow equations for the dimensionless couplings Eqs.~\eqref{betag1t}-\eqref{betag4t}, i.e. solutions $g^*=(g_1^*,g_2^*,g_3^*,g_4^*)$ of
\begin{align}
\beta_{g_1}=\partial_t \hat g_1&=0\\
\beta_{g_2}=\partial_t \hat g_2&=0\\
\beta_{g_3}=\partial_t \hat g_3&=0\\
\beta_{g_4}=\partial_t \hat g_4&=0\,.
\end{align}
In general, there are several fixed point solutions and we are interested in their stability, i.e. if they are reachable without fine-tuning. The stability of a fixed point can be determined by calculating the eigenvalues of the stability matrix evaluated at the fixed point
\begin{equation}
\left.\frac{\partial}{\partial_{g_i}}\beta_{g_j}\right|_{g^*}
\end{equation}
The fixed point is stable, when all eigenvalues are negative. A negative eigenvalue corresponds to an irrelevant direction and a positive eigenvalue to a relevant one.

In general, the existence of a real solution and the number of relevant directions depends on $d_0$ and $N$. However, we find that all fixed-point solutions possess one or more relevant directions, i.e. all are unstable.
Among others, we can identify the interacting fixed point found in Ref.~\onlinecite{PhysRevResearch.1.033206}, where just one HOVH point was considered, so that only $g_4$ is present. In our more general setup with several HOVH points, it is given by
\begin{align}
g_1^*=g_2^*=g_3^*=0\,\quad\text{and}\quad
g_4^*=\frac{1}{1-d_0}\kappa\,.
\end{align}
It has two relevant directions for $d_0<1/3$ and one relevant direction for $d_0>1/3$.
Furthermore, there are two more possible solutions with just one relevant direction for sufficiently small $d_0$.

For $N\!=\!2$ and $d_0\geq 2(N\!-\!1)/(N^2\!+\!2(N\!-\!1))$, they are
\begin{align}
g_1^*&=g_3^*=0\,,\nonumber\\
g_2^*&=\frac{(N-2)(1-2d_0)}{2d_0[(N-2)^2-d_0(N^2-6(N-1))]}\kappa\nonumber\\
&\quad\mp\frac{\sqrt{d_0[d_0(N^2+2(N-1))-2(N-1)]}}{2d_0[(N-2)^2-d_0(N^2-6(N-1))]}\kappa\,,\nonumber\\
g_4^*&=\frac{d_0(2+N(N-2))}
{2d_0[(N-2)^2-d_0(N^2-6(N-1))]}\kappa\nonumber\\
&\quad \pm\frac{(N-2)\sqrt{d_0[d_0(N^2+2(N-1))-2(N-1)]}}
{2d_0[(N-2)^2-d_0(N^2-6(N-1))]}\kappa\,.\nonumber
\end{align}
For $N=3$, they have two relevant directions.
The two solutions with one relevant direction for $N=3$ are
\begin{align}
g_1^*&=\frac{9+16d_0-\sqrt{d_0(72+481d_0+72d_0^2)}}{9d_0(d_0-1)}\kappa\,,\nonumber\\
g_2^*&=-2g_4^*=\frac{25d_0-\sqrt{d_0(72+481d_0+72d_0^2)}}{9d_0(d_0-1)}\kappa\,,\nonumber\\
g_3^*&=\frac{16d_0+9d_0^2-\sqrt{d_0(72+481d_0+72d_0^2)}}{9d_0(d_0-1)}\kappa\,,
\end{align}
and for $d_0<9/13$
\begin{align}
g_1^*&=g_2^*=\frac{9+8d_0+\sqrt{d_0(108+213d_0+104d_0^2)}}{d_0(13d_0-9)}\kappa\,,\nonumber\\
g_3^*&=\frac{12d_0+13d_0^2+\sqrt{d_0(108+213d_0+104d_0^2)}}{d_0(13d_0-9)}\kappa\,,\nonumber\\
g_4^*&=-\frac{21d_0+\sqrt{d_0(108+213d_0+104d_0^2)}}{2d_0(13d_0-9)}\kappa\,.
\end{align}
%

\section{Susceptibility exponents}
\label{sec:exponents}

To determine the leading ordering tendency, we compare the susceptibilities for superconducting, charge and spin orders. The largest susceptibility has the largest exponent, cf. Eqs.~\eqref{exps}-\eqref{expsc}. Furthermore, the exponent has to be larger than 1/2 for the susceptibility to diverge approaching the critical scale. Based on the largest exponent, we assign the ordering tendency to the fixed trajectories. We plot the different exponents as function of $d_0$ for the stable fixed trajectories in Fig.~\ref{fig:exponents}.

\end{appendix}

\bibliography{Bib_HOVH}

\end{document}